\documentclass[11pt]{article}
\usepackage{graphics}
\begin{document}
\title{\normalsize \hspace*{10.7cm}LNF-00/026(P)\\
                   \hspace*{10.2cm}UWThPh-2000-40\\
\vspace*{2.0cm}
\Large \textbf{\boldmath $K\rightarrow \pi\pi e^+e^-$ decays and chiral
low-energy constants*}}
\author{Hannes Pichl\\ \\
\textit{INFN, Laboratori Nazionali di Frascati,}\\
\textit{P.O. Box 13, I-00044 Frascati, Italy} \\
and \\
\textit{Institut f\"ur Theoretische Physik, Universit\"at Wien,}\\
\textit{ A-1090 Wien, Austria}}
\date{}
\maketitle
\thispagestyle{empty}
\begin{abstract}
\noindent
The branching ratios of the measured decay 
$K_L\rightarrow \pi^+\pi^-e^+e^-$ and of the still unmeasured decay 
$K^+\rightarrow\pi^+\pi^0e^+e^-$ are calculated to next-to-leading order in 
Chiral Perturbation Theory (CHPT). Recent experimental results are 
used to determine two possible values of the combination $(N^r_{16}-N_{17})$ 
of weak low-energy couplings (LECs) from the $\mathcal{O}(p^4)$ chiral
Lagrangian. The obtained values are compared to the predictions of 
theoretical approaches to weak counter\-term couplings to distinguish
between the two values. Using the favoured value of the combination
$(N^r_{16}-N_{17})$ and taking into account additional assump\-tions 
suggested by the considered models, one obtains the branching ratio of the 
se\-cond decay as a function of the unknown combination 
$(N^r_{14}+2N^r_{15})$ of weak low-energy couplings. Finally, using 
values of the individual LECs derived from a particular model, one 
predicts the branching ratio of the $K^+$ decay.
\end{abstract}
\vspace*{1.0cm}
\nopagebreak[4]
\begin{center}
*Work supported in part by TMR, EC-Contract No. ERBFMRX-CT980169 \\
(EURODA$\Phi$NE) 
\end{center}
\newpage
\setcounter{page}{1}
\section{Introduction}
During the last years, there has been a lot of theoretical and experimental 
interest in the decay of the $K_L$ into a pair of charged pions and a pair of
leptons. This interest focused on the decay width itself
\cite{SW,HS,ESW,SAV,KTEV1,TAK,MAZ,BIZZ} and on the possibility of constructing 
CP-violating observables \cite{SW,HS,ESW,SAV,MAZ,BIZZ,ESWW,ECKPI,KTEV2,KTEV3} 
as well as on other related topics \cite{LS1,LS2}. 
\newline
\hspace*{0.5cm}
From the experimental analysis of the corresponding radiative decay, it was 
found that the decay amplitude consists of a bremsstrahlung component and a
direct emission part. The contribution due to bremsstrahlung is given via 
Low's theorem by the amplitude of the decay $K_L\rightarrow
\pi^+\pi^-$. This amplitude is mainly due to the $K^0_1$ admixture, which 
allows for this decay (indirect CP violation). As a consequence, the final
state of the radiative decay can be found to be in CP-even as well as
CP-odd configurations. Hence, in principle, there is interference between 
the CP-conserving parts of the direct emission amplitude and the
CP-violating bremsstrahlung amplitude. But as long as the polarization of
the on-shell photon is not measured, this interference is not accessible.
This is the reason why one looks directly to the decay with a lepton pair, 
since the angle between the two planes spanned by the pions and leptons 
can be used to construct a CP-violating observable 
\cite{SW,HS,ESW,SAV,ESWW,ECKPI}.   
\newline
\hspace*{0.5cm}
In this paper, I do not focus on the CP-violating aspects of
this decay. I calculate the decay amplitude in CHPT up to
$\mathcal{O}(p^4)$ and use the most recent available data from 
experiments \cite{KTEV2,BAR,KETT} (which were mostly dedicated to the study of 
possible CP-violating effects) to derive a value for the unknown 
combination $(N_{16}^r(\mu)-N_{17})$ of low-energy couplings (LECs) from
the weak $\mathcal{O}(p^4)$ chiral Lagrangian. Until 
now, theoretical predictions can only be compared to the branching ratio 
over the entire phase space, which makes it impossible to extract a precise 
value for this combination. Furthermore, it is not possible to determine this 
value unambiguously from experiment, therefore one has to turn to
LEC models and their predictions for low-energy couplings to find the 
favoured value.
\newline\hspace*{0.5cm}
Once the value of this particular combination is fixed, I use it as 
input together with additional assumptions about the weak LEC $N_{17}$
for the second non-leptonic decay discussed in this paper: 
$K^+\rightarrow\pi^+\pi^0e^+e^-$. If we use new data from the
corresponding radiative decay $K^+\rightarrow \pi^+\pi^0\gamma$
\cite{E787}, we can give the magnetic amplitude of $K^+\rightarrow \pi^+\pi^0
e^+e^-$ without any unknown para\-meter at $\mathcal{O}(p^4)$ and it is 
possible to predict the branching ratio $BR(K^+\rightarrow \pi^+\pi^0
e^+e^-)$ as a function of the unknown combination $(N^r_{14}(\mu)+
2N^r_{15}(\mu))$ of weak LECs.
\section{Effective chiral Lagrangians}
Chiral Perturbation Theory \cite{GL1,GL2} is the ideally suited 
framework to discuss these processes. It is the low-energy realization of 
the Standard Model respecting the approximate chiral symmetry of the 
light quark sector. In fact, the demand of invariance under chiral rotations 
(in our case, these are $SU(3)$ rotations) allows one to write 
down the most general effective Lagrangian of strong interactions amongst 
the light pseudoscalar meson octet. 
The approximate chiral symmetry $SU(3)_L\times SU(3)_R$ seems to be realized
\`a la Nambu-Goldstone, which means that it is spontaneously broken to the
well-known $SU(3)_V$. The breakdown of the symmetry gives rise to eight
almost massless would-be Goldstone bosons because there are eight broken
axial generators. According to Goldstone's theorem, the quantum
numbers of these particles are fixed by the quantum numbers of the
broken generators, thus one identifies the light pseudoscalars with these
particles. 
\newline\hspace*{0.5cm}
In the scheme of Gasser and Leutwyler \cite{GL2}, the most general
$\mathcal{O}(p^2)$ Lagrangian including strong, electromagnetic and
semileptonic weak interactions reads as follows:
\begin{equation}
\label{L1}
\mathcal{L}_2=\frac{F^2}{4}\langle D_{\mu}UD^{\mu}U^{\dagger}+\chi
U^{\dagger}+\chi^{\dagger}U\rangle,
\end{equation}
where $D_{\mu}U$ is the covariant derivative with respect to external,
non-propagating fields. If we specialize to the case of external photons, 
\begin{eqnarray}
\label{CovDer}
D_{\mu}U&=&\partial_{\mu}U+ieA_{\mu}[Q,U], \nonumber \\
Q&=&\frac{1}{3}\cdot\mbox{diag}(2,-1,-1),
\end{eqnarray}
where Q is the quark charge matrix for the flavours $up$, $down$, $strange$.
$U$ is a $3\times 3$ unitary matrix which has to be expanded to the relevant 
order in $\Phi$:     
\begin{equation}
U(\Phi) = e^{i\sqrt{2} \Phi /F} ,
\end{equation}
where the mesons are collected in the matrix $\Phi$:
\begin{eqnarray}
\Phi&=&\left( \begin{array}{ccc}
\frac{\pi^0}{\sqrt{2}}+\frac{\eta_8}{\sqrt{6}} & \pi^+ & K^+ \\
\pi^- & - \frac{\pi^0}{\sqrt{2}} + \frac{\eta_8}{\sqrt{6}} & K^0 \\
K^- & \bar{K}^0 & -2\frac{\eta_8}{\sqrt{6}}
\end{array} \right).
\end{eqnarray}
$F$ equals to lowest order the pion decay constant, $F_{\pi}=92.4$ MeV.
In general, $\chi$ contains external scalar and pseudoscalar matrix-valued
fields, but here it is proportional to the quark mass matrix. In this way,
explicit chiral symmetry breaking can be incorporated in the effective
Lagrangians in a very elegant way:
\begin{eqnarray}
\chi = 2B_0\cdot\mbox{diag}(m_u,m_d,m_s).
\end{eqnarray}
$B_0$ is related to the order parameter of the spontaneous breakdown of the
chiral symmetry, the quark condensate. It will not appear explicitly
because it can be absorbed in the squared meson masses.
\newline\hspace*{0.5cm}
For the calculation of non-leptonic kaon decays, we also need an effective
Lagrangian describing the weak interactions of the mesons. This
effective weak Lagrangian cannot be invariant under chiral rotations, hence
chiral invariance cannot be the guideline. Starting from an
effective strangeness-changing $\Delta S=1$ four-quark Hamiltonian, one 
writes down a hadronically realized Lagrangian that transforms in the same 
way under $SU(3)_L\times SU(3)_R$ as this Hamiltonian. At lowest order, the 
needed weak Lagrangian is found to be:   
\begin{eqnarray}
\label{L2}
\mathcal{L}_2^{\Delta S=1}&=&G_8\langle \lambda L_{\mu}L^{\mu}\rangle
+G_{27} \big [ L_{\mu23}L^{\mu}_{11}+\frac{2}{3}L_{\mu 21}L^{\mu}_{13}
\big ] + h.c.,
\end{eqnarray}
where $\lambda=(\lambda_6-i\lambda_7)/2$ projects out the correct octet quantum
numbers and $L_{\mu}=iF^2 U^{\dagger}D_{\mu}U$ is the hadronic left-chiral
current in analogy to the left-chiral quark current at the level of the
effective Hamiltonian. The two couplings $G_8$ and $G_{27}$ have to be
obtained from experiment and the determination of these couplings involves
some subtleties. In principle, the couplings are obtained from
$K\rightarrow \pi\pi$ decays. Comparison of experiments with the leading order 
$\mathcal{O}(p^2)$ calculations yields the 'canonical' values 
$|G_8|\simeq 9.1\cdot 10^{-6}\mbox{ GeV}^{-2} \; \mbox{and} 
\; G_{27}/G_8\simeq 1/18$, where this approximate ratio of the two couplings 
introduces uncertainties when the $27$-plet coupling enters the game. 
However, due to the smallness of the $27$-plet coupling, one
can usually neglect this part of the Lagrangian unless the octet 
contribution vanishes. Then also the $27$-plet contribution may become  
important (see Sect. 3.2). 
\newline\hspace*{0.5cm}
For completeness, one should remark that in 
Ref. \cite{KMW1} the relevant $K\rightarrow \pi\pi$ decays were analyzed up 
to $\mathcal{O}(p^4)$ and it was found that these additional corrections 
contribute to $G_8$ with about 30\%, whereas the $G_{27}$ coupling is only 
modified by a few percent. Thus, if one takes into account these order
$p^4$ corrections, the value of the coupling $|G_8|$ appearing in (\ref{L2})
should better be $\sim 6.4\cdot 10^{-6}\mbox{ GeV}^{-2}$. Throughout this
work, however, I am using the canonical standard values.  
\newline
\hspace*{0.5cm}
The chiral Lagrangians (\ref{L1}) and (\ref{L2}) allow us to calculate tree-level 
amplitudes of chiral order $p^2$ and one-loop diagrams of chiral 
order $p^4$ which usually introduce divergences. In order to get rid of
these divergences and to take into account further finite local 
corrections appearing at $\mathcal{O}(p^4)$, e.g. through new interactions 
arising from the chiral anomaly, one also has to consider the most general 
$\mathcal{O}(p^4)$ interaction Lagrangians.
\newline\hspace*{0.5cm}
The most general strong Lagrangian of order $p^4$, invariant under C, P 
and chiral transformations, was again given by Gasser and Leutwyler \cite{GL2}.
There is only one term in this Lagrangian that contributes to the final 
results in this work: 
\begin{equation}
\label{CTs}
\mathcal{L}_4 = -iL_9\langle F_R^{\mu\nu}D_{\mu}UD_{\nu}U^{\dagger}+
F_L^{\mu\nu}D_{\mu}U^{\dagger}D_{\nu}U\rangle .
\end{equation}
Since we are interested in external photons, the $F^{\mu \nu}_{L,R}$
tensors are proportional to the ordinary electromagnetic field strength tensor:
\begin{equation}
\label{Ten}
F^{\mu \nu}_{L}=-eQF^{\mu \nu}=F^{\mu \nu}_R, \quad F^{\mu\nu}=\partial^{\mu}
A^{\nu}-\partial^{\nu}A^{\mu}.
\end{equation}
Every new term in the strong Lagrangian of order $p^4$ is furnished with
an a priori unknown low-energy coupling (LEC) \cite{GL2}. 
Since all divergences appear as local polynomials, one can absorb the 
divergences of the loop amplitude in the LECs. The general structure of a 
LEC reads as
\begin{eqnarray}
\label{LEC}
L_i=L^r_i(\mu)+\Gamma_i\Lambda(\mu), &&
\Lambda(\mu)=\frac{\mu^{d-4}}{16\pi^2} 
\Big [\frac{1}{d-4}-\frac{1}{2}(\ln (4\pi)+1-\gamma_E)\Big ],
\end{eqnarray} 
where $\gamma_{E}=0.5772157$ is the Euler-Mascheroni constant.
This is also true for weak LECs $N_i$. The coefficients $\Gamma_i$ arise 
from the one-loop generating functional. Because of the 
regularization procedure, the measurable couplings $L_i^r(\mu)$
(and $N_i^r(\mu)$) become scale dependent. In the end, this
scale dependence must be compensated by the scale dependent parts of loop diagrams.
One should also note that the chiral subtraction prescription differs from the
usual modified MS prescription. 
\newline\hspace*{0.5cm}
The new octet weak interactions are organized like this \cite{KMW2,EKW}:
\begin{equation}
\label{L4}
\mathcal{L}_4^{\Delta S=1}=G_8 F^2 \sum_i N_i W_i+h.c.
\end{equation} 
For the non-leptonic kaon decays under consideration, only the operators
$W_{14},W_{15}$, $W_{16},W_{17}$ and $W_{28}$, $W_{29},W_{30},W_{31}$ 
contribute; they are listed explicitly:
\begin{eqnarray}
\label{LEle}
W_{14} & = & i\langle \lambda\{F_L^{\mu\nu}+U^{\dagger}F_R^{\mu\nu}U,D_{\mu}U^{\dagger}
D_{\nu}U\}\rangle, \nonumber\\
W_{15} & = & i\langle \lambda
D_{\mu}U^{\dagger}(UF_L^{\mu\nu}U^{\dagger}+F_R^{\mu\nu})D_{\nu}U\rangle,
\nonumber \\
W_{16} & = & i\langle \lambda\{F_L^{\mu\nu}-U^{\dagger}F_R^{\mu\nu}U,D_{\mu}U^{\dagger}
D_{\nu}U\}\rangle, \nonumber \\
W_{17} & = & i\langle \lambda D_{\mu}U^{\dagger}(U F_{L}^{\mu\nu}U^{\dagger}-F_R^{\mu\nu})
D_{\nu}U\rangle. 
\end{eqnarray}
The magnetic terms (proportional to $\epsilon_{\mu\nu\rho\sigma}$) are given by
\begin{eqnarray}
\label{LMag}
W_{28} & = & i\epsilon_{\mu \nu \rho \sigma}\langle \lambda
D^{\mu}U^{\dagger}U\rangle \langle
U^{\dagger}D^{\nu}UD^{\rho}U^{\dagger}D^{\sigma}U\rangle, \nonumber \\ 
W_{29} & = & 2 \langle \lambda[U^{\dagger}\tilde{F}_R^{\mu\nu}U,D_{\mu}U^{\dagger}
D_{\nu}U]\rangle, \nonumber \\
W_{30} & = & \langle \lambda
U^{\dagger}D_{\mu}U\rangle\langle(\tilde{F}_L^{\mu\nu}+U^{\dagger}\tilde{F}_R^{\mu\nu}U)
D_{\nu}U^{\dagger}U\rangle, \nonumber \\
W_{31} & = & \langle \lambda U^{\dagger}D_{\mu}U\rangle
\langle(\tilde{F}_L^{\mu\nu}-U^{\dagger}\tilde{F}_R^{\mu\nu}U)D_{\nu}U^{\dagger}U\rangle,
\end{eqnarray}       
with $\tilde{F}^{\mu\nu}_{L,R}$ the dual tensor of (\ref{Ten}), 
$\tilde{F}^{\mu\nu}_{L,R}=\epsilon^{\mu\nu\rho\sigma}F_{\rho\sigma L,R}$.
\newline\hspace*{0.5cm}
Finally, we introduce a Lagrangian that embodies contributions from 
reducible diagrams with a WZW-vertex and an $\mathcal{O}(p^2)$ $\Delta S=1$ 
vertex. It only contributes to the $K^+$ decay and is given by 
\cite{ENP1,BIJ,ENP2}
\begin{equation}
\label{LWZW}
\mathcal{L}^{\Delta S=1}_{an}=\frac{ieG_8}{8\pi^2F}\tilde{F}^{\mu\nu}\partial_{\mu}
\pi^0K^+\stackrel{\leftrightarrow}{D}_{\nu}\pi^-,
\end{equation}
where $\tilde{F}^{\mu\nu}$ is the dual of the ordinary electromagnetic field
strength tensor (\ref{Ten}), $\tilde{F}^{\mu\nu}=\epsilon^{\mu\nu\rho\sigma}
F_{\rho\sigma}$, and the covariant derivative is the usual QED derivative.
\section{Amplitudes}
For both decays, the general form of the invariant amplitude due to 
covariance \nolinebreak[4]is 
\begin{equation} 
\label{AMP}
\mathcal{A}=\frac{e}{q^2}V_{\mu}\bar{u}(k_-)\gamma^{\mu}v(k_+),
\end{equation}
where $q=k_-+k_+$ is the momentum of the virtual photon, $k_-$ and $k_+$
are the momenta of the electron and positron, respectively. $iV_{\mu}$ is
the generic weak $K\pi\pi(\gamma^{\star})$ vertex, calculated in CHPT. It
is decomposed in an electric and a magnetic part
\begin{equation}
V_{\mu}=\mathcal{F}_1 p_{1\mu}+\mathcal{F}_2
p_{2\mu}+\mathcal{M}\epsilon_{\mu\nu\rho\sigma}p_1^{\nu}p_2^{\rho}q^{\sigma},
\end{equation}
where $p_1$ and $p_2$ are the outgoing momenta of the $\pi^+$ and  
$\pi^-(\pi^0)$ and $\mathcal{F}_1$, $\mathcal{F}_2$, $\mathcal{M}$ are form
factors containing the dynamics of
the two processes. A separate term proportional to the photon momentum vanishes
because of the Dirac equation. The form factors are either constants or scalar
functions of various products of the involved momenta.
\subsection{\boldmath $K_L\rightarrow\pi^+\pi^-\gamma^{\star}$ Amplitudes}
\begin{figure}[h]
\resizebox{1.0\textwidth}{!}{
\includegraphics{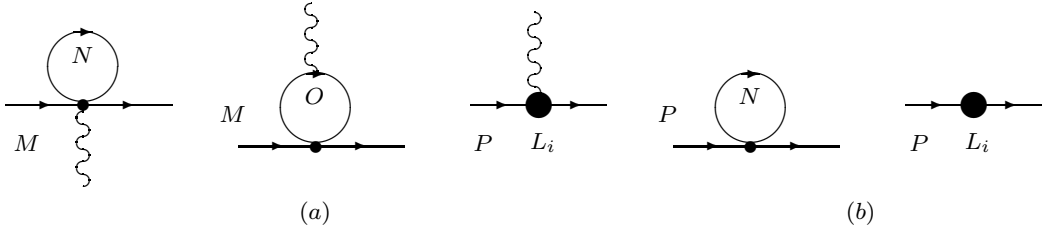}}
\caption{Strong insertions. a) Loops, where the photon is emitted at 
the vertex or by a charged meson in the loop, and the generic photon 
emitting counterterm proportional to $L_i$. $M$ denotes $K^0_2$, $\pi^+$ or 
$\pi^-$. $N$ denotes the allowed particles in the loop: $\pi^0$, $\pi^+$, 
$K^+$, $\eta_8$, $K^0_1$ or $K^0_2$. $O$ denotes any charged pseudoscalar. 
$P$ denotes $\pi^+$ or $\pi^-$. b) Generic loop and generic counterterm 
vertex proportional to $L_i$ without a photon.}
\label{f2}
\end{figure}

\begin{figure}[h]
\hspace*{0.1cm}
\resizebox{0.95\textwidth}{!}{
\includegraphics{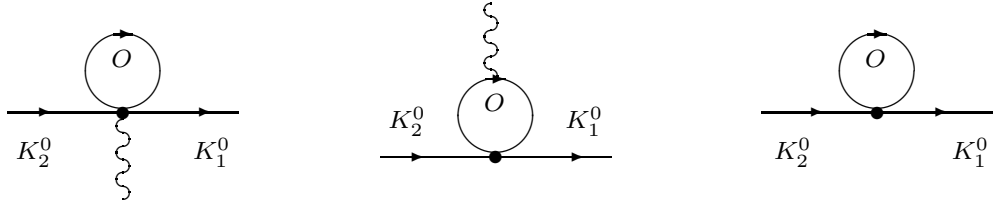}}
\caption{Strong loop insertions for the external $K^0_2$ line
that generate a $K^0_1$ propagator. The contributions derived from these 
insertions, however, are finite. $O$ denotes a charged pion or kaon.} 
\label{f3}
\end{figure}

\begin{figure}[h]
\hspace*{3.2cm}
\resizebox{0.55\textwidth}{!}{\includegraphics{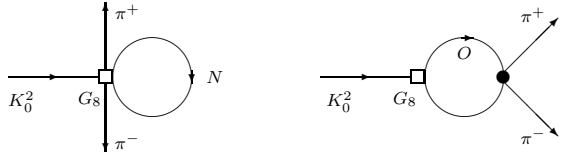}}
\caption{Weak loop diagrams: the basic tadpole diagram (left) and 
the basic diagram of topology 1 (right). $N$ and $O$ denote the same particles as 
in Figure \ref{f2}.}
\label{f4}
\end{figure}

\begin{figure}[h]
\hspace*{3.2cm}
\resizebox{0.55\textwidth}{!}{\includegraphics{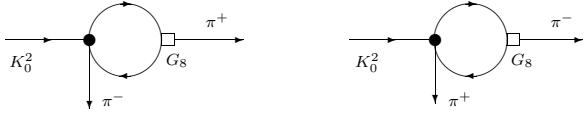}}
\caption{Weak loop diagrams: the basic diagram of topology
    2 (left), the basic diagram of topology 3 (right).} 
\label{f5}
\end{figure}

This decay had already been considered in the framework of CHPT in
\cite {ESW}. The authors of Ref. \cite{ESW} used a different basis of 
counterterms (this change of basis is only valid as long as one is only 
interested in photons in (\ref{CovDer})) and a different approximation of the magnetic part
of the amplitude not taking into account any energy dependence. 
The present calculation considers this energy dependent part \cite {KTEV2}, 
too, and additionally serves as a check on the results in Ref. \cite{ESW}. 
\newline\hspace*{0.5cm}
For this decay, I assume strong isospin 
conservation, i.e. the $up$ and the $down$ quark have equal masses. Hence, 
the Gell-Mann-Okubo mass relation holds and we only have to deal with two 
independent masses: $3m_{\eta_8}^2=4m_{K}^2-m_{\pi}^2$. It will be used to
simplify parts of the one-loop amplitude given in the appendix.
\newline\hspace*{0.5cm}
In this paper, I use the following definitions: $K_L=K^0_2+
\epsilon K^0_1$, where $CP|K^0_1\rangle =+|K^0_1\rangle$ and 
$CP|K^0_2\rangle = -|K^0_2\rangle$. $K^0_1$ and $K^0_2$ are related
to the strangeness eigenstates $K^0$ and $\bar{K}^0$ through the following 
expressions:
\begin{eqnarray}
K^0_1 = \frac{1}{\sqrt{2}}(K^0-\bar{K}^0), &&
K^0_2 = \frac{1}{\sqrt{2}}(K^0+\bar{K}^0).
\end{eqnarray}
\hspace*{0.5cm}The tree-level amplitude is entirely due to the $K^0_1$ 
admixture, since we do not consider direct sources of 
CP violation. In any case, the tree-level contribution is rather 
suppressed, especially when compared to the $K^+$ decay (Sect. 3.2).
\begin{figure}
\hspace*{1.2cm}
\resizebox{0.85\textwidth}{!}{
\includegraphics{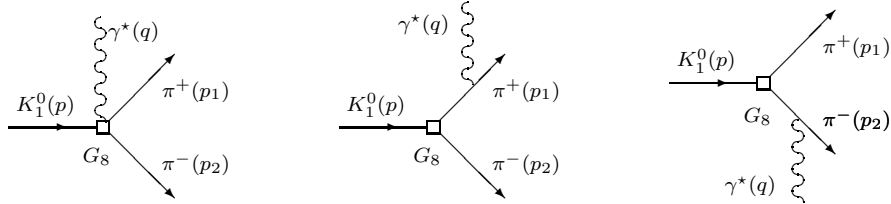}}
\vspace*{0.1cm}
\caption{Tree-level diagrams for $K_L\rightarrow\pi^+\pi^-\gamma^{\star}$. 
At tree level, the $K_L$ transition is entirely due to $K^0_1$ admixture.}
\label{f1}
\end{figure}
From Figure \ref{f1} one obtains the following tree-level form factors: 
\begin{eqnarray}
\label{ALtree}
\mathcal{F}^{Lt}_1=-i\epsilon\frac{4eG_8F}{2qp_1+q^2}(m_K^2-m_{\pi}^2),
&&
\mathcal{F}^{Lt}_2=i\epsilon\frac{4eG_8F}{2qp_2+q^2}(m_K^2-m_{\pi}^2),
\end{eqnarray}
where $\epsilon\simeq 2.27\cdot 10^{-3}e^{i44^o}$ is the parameter of indirect 
CP violation. In the remainder of the paper, we do not take into account 
$\mathcal{O}(p^4)$ corrections proportional to $\epsilon$ to the electric 
form factors. As mentioned in Sect. 2, a value of $|G_8|\simeq 9.1\cdot 
10^{-6}\mbox{ GeV}^{-2}$ already amounts to some $\mathcal{O}(p^4)$ 
contributions.  
\newline
\hspace*{0.5cm}
The magnetic form factor can only arise through the four weak counterterms
$W_{28}$, \ldots,$W_{31}$ and it is in fact a result of the chiral anomaly. 
It is necessarily finite and does not have any energy dependence at this
order. CHPT generates the following direct emission magnetic form factor:
\begin{eqnarray}
\label{M1}
\mathcal{M}^L&=&\frac{-16eG_8}{F}(N_{29}+N_{31})\nonumber \\
&=&\frac{-eG_8}{2\pi^2F}(a_2+2a_4),
\end{eqnarray}
where I have used the 'magnetic' notation of Refs. \cite{BIJ,ENP2}. 
These magnetic LECs are also still unknown. Experiments exhibit a large 
sensitivity of the magnetic amplitude to the energy of the 
emitted photon, therefore I will use the experimental results of 
\cite{KTEV2} (rather than the old results of the experiment by Ramberg et 
al., \cite{Ram}) to estimate the magnetic contribution. The authors of
Ref. \cite{KTEV2} use the papers by Sehgal et al. \cite{SW,HS} as the theoretical background to model
their Monte Carlo, but additionally introduce an energy dependence in the magnetic 
amplitude through a form factor that involves a kind of a $\rho$ propagator:
\begin{eqnarray}
\label{AmagKTeV}
\mathcal{M}^L=e|f_s|\frac{\tilde{g}_{M1}}{m_K^4}\mathcal{W}, &&
\mathcal{W}=\Big [1+\frac{a_1/a_2}{(m_{\rho}^2-m_K^2)+2m_K 
E^{\star}_{\gamma}}\Big ].
\end{eqnarray}
Ansatz (\ref{AmagKTeV}) cannot be compared directly to the magnetic form 
factor in \cite{SW,HS,ESW,SAV,ESWW}. Consequently accor\-ding to 
\cite{KTEV2}, one should identify the average of
$\tilde{g}_{M1}\mathcal{W}$ over the allowed range of $E^{\star}_{\gamma}$, 
the energy of the virtual photon, with the original magnetic coupling used
in Refs. \cite{SW,HS,ESW,SAV,ESWW}. $|f_s|\simeq 3.9\cdot 10^{-4}\;\mbox{MeV}$ is the absolute value of the decay amplitude of
$K_S\rightarrow\pi^+\pi^-$, and the experiment \cite {KTEV2} gave for
the magnetic coupling $|\tilde{g}_{M1}|=1.35_{-0.17}^{+0.20}\mbox{(stat.)}\pm 0.04
\mbox{(syst.)}$ and for $a_1/a_2=-0.720\pm 0.028
\mbox{(stat.)}\pm 0.009\mbox{(syst.)}\;\mbox{GeV}^2$. 
($a_2$ in the fraction above is not the same as the LEC $a_2$ 
in (\ref{M1}).) These numbers were obtained from the entire KTeV 1997 
data set of more than 1811 events above background \cite{KTEV2}. In fact, it is also 
this data set and this parametrization that were used to extract the most 
recent value of the branching ratio of $K_L\rightarrow \pi^+\pi^-e^+e^-$
\cite{BAR,KETT}. Additionally, the fraction $a_1/a_2$ was found  
from the corresponding radiative decay $K_L\rightarrow \pi^+\pi^-\gamma$
to be $-0.729\pm 0.026 \mbox{(stat.)}\pm 0.015\mbox{(syst.)}\;\mbox{GeV}^2$
\cite{KTEV3}, which is clearly in perfect agreement. The errors of these 
quantities are the sources of by far the most important contributions to 
the uncertainties in the extraction of the LEC combination
$(N^r_{16}(\mu)-N_{17})$.
\newline
\hspace*{0.5cm}
The electric form factors at $\mathcal{O}(p^4)$ show the pleasant feature
that one can obtain the form factor $\mathcal{F}^L_2$ from the expression for
$\mathcal{F}^L_1$ by simply exchanging the pion momenta $p_1$ and $p_2$. 
At this order, there is no change of sign as at the tree 
level (\ref{ALtree}), since we concentrate on the CP-conserving part of the decay. 
\newline\hspace*{0.5cm}
We begin the discussion of next-to-leading-order electric form factors by
considering strong loops and strong counterterm contributions. The starting 
point for our analysis is the collection of diagrams in Figure \ref{f1}, 
where one replaces $K^0_1$ with $K^0_2$. Wave function renormalization
graphs will be neglected, since the tree-level amplitude of 
$K^0_2\rightarrow\pi^+\pi^-\gamma^{\star}$ vanishes. 
\newline\hspace*{0.5cm}
Removing the photon in the left diagram of Figure \ref{f1} and replacing
the external kaon line with the appropriate loop diagrams from Figure 
\ref{f2}a, one obtains diagrams that are found to vanish. This feature is
due to the structure of the weak $K^0_2\pi^+\pi^-$ vertex. Appropriate 
replacement of the kaon line in Figure \ref{f1} with the loop diagrams
drawn in Figure \ref{f3}, however, yields
a finite $K^0_1$ propagator contribution to the electric form factors. The
contribution must be finite, since there are no counterterms to compensate
a divergence. 
This kind of diagrams was already considered in \cite{SW,HS} and the sum of 
the diagrams yields
\begin{eqnarray}
\mathcal{F}_{11}^{Ll}&=&\frac{-ieG_8}{(d-1)F}\frac{1}{[(p_1+p_2)^2-m_K^2]}
(m_{\pi}^2+2p_1p_2)\nonumber \\
&&\hspace*{-0.0cm}\Big\{B(q^2,m_{\pi}^2,m_{\pi}^2)(4m_{\pi}^2-q^2)+(4-2d)A(m_{\pi}^2)\nonumber \\
&&\hspace*{-0.15cm}-B(q^2,m_K^2,m_K^2)(4m_K^2-q^2)-(4-2d)A(m_K^2)\Big\}.
\end{eqnarray} 
\hspace*{0.5cm}
A second kind of strong $\mathcal{O}(p^4)$ corrections is obtained from the
two remaining bremsstrahlung graphs in Figure \ref{f1} in two ways: first, 
by either putting the loop diagram (without photon) of Figure \ref{f2}b 
or the counter\-term insertion (without photon) of Figure \ref{f2}b instead 
of the internal pion lines. Secondly, by replacing the scalar QED vertices 
with the diagrams of Figure \ref{f2}a. Focusing on the 
counterterms first, one finds that counterterms proportional to the LECs 
$L_4$, $L_5$ and $L_9$ from the order $p^4$ Lagrangian \cite{GL2} are
allowed to contribute. Calculating the sum of all the diagrams of the
second kind, however, one discovers that only the contribution from the
counterterm (\ref{CTs}) proportional to $L_9$ survives. The final
correction of strong order $p^4$ diagrams of the second kind to the form 
factors is given by a very condensed and compact result:  
\begin{eqnarray}
\label{FFLl12}
\mathcal{F}_{12}^{Ll} & = & \frac{-ieG_8}{F}
\Big\{-4q^2 L_9-\frac{d-2}{d-1}\Big [2A(m_{\pi}^2)+A(m_K^2)\Big ]\nonumber
\\&&+\frac{1}{d-1}\Big [(4m_{\pi}^2-q^2)B(q^2,m_{\pi}^2,m_{\pi}^2)\nonumber \\
&&+\frac{1}{2}(4m_K^2-q^2)B(q^2,m_K^2,m_K^2)\Big ]\Big \}.
\end{eqnarray}
\noindent 
The tadpole integral $A(m^2)$ and $B(p^2,m^2,m^2)$, the scalar
two-propagator integral, are defined in Appendix A. $d$ is the spacetime 
dimension coming from dimensional regularization. 
\newline\hspace*{0.5cm}
Actually, the form factor 
$F^{Ll}_{12}$ contains divergences stemming from the functions $A$ and $B$
which are removed by the divergent part of the strong counter\-term
coupling $L_9$ (coming from Figure \ref{f2}a). In the finite amplitude with 
strong $\mathcal{O}(p^4)$ insertions, the measurable part of 
$L_9$ shows up: $L_9^r(m_{\rho})=(6.9\pm 0.7)\cdot 10^{-3}$ \cite{GL2}. 
Throughout this paper, I always choose $\mu=m_{\rho}$ as renormalization 
scale.  
\newline
\hspace*{0.5cm}
Weak counterterms only contribute through a diagram obtained from
the direct emission diagram in Figure \ref{f1} by replacing $K^0_1$ with 
$K^0_2$ and putting in the counterterm vertex from (\ref{L4}) and 
(\ref{LEle}) instead of the lowest-order vertex. All occuring divergences from weak loop diagrams must be
removed by this local counterterm contribution. The weak counterterms 
produce the following contribution to the electric form factors:
\begin{equation}
\label{FFLl13}
\mathcal{F}_{13}^{Ll}=\frac{2ieG_8}{3F}q^2[N_{14}-N_{15}-3(N_{16}-
N_{17})],
\end{equation}
where all renormalized LECs $N^r_i$ (compare with (\ref{LEC})) depend on a 
scale $\mu=m_{\rho}$. 
Since the coefficient $\Gamma_{17}$ is found to vanish, the LEC $N_{17}$
is independent of the renormali\-zation scale. The renormalized parts 
of the low-energy couplings enter into the amplitude of the decay,
therefore it is important to know their finite values. 
\newline\hspace*{0.5cm}
The combination $(N_{14}^r-N_{15}^r)$ also appears in the 
counterterm part of the form factor describing the decay 
$K^+\rightarrow \pi^+e^+e^-$ within the expression \cite{EPR}
\begin{eqnarray}
\label{W+}
w_+=\frac{64\pi^2}{3}[N^r_{14}(\mu)-N^r_{15}(\mu)+3L^r_9(\mu)]+\frac{1}{3}\ln\Big [\frac{\mu^2}{m_K m_{\pi}}\Big ].
\end{eqnarray}
Old experiments \cite{ALL} fixed $w_+$ to be $0.89^{+0.24}_{-0.14}$, which
corresponds to a value of $(N^r_{14}(m_{\rho})-N^r_{15}(m_{\rho}))
\simeq -0.02$. A more refined theoretical analysis of this decay also took 
into account $\mathcal{O}(p^6)$ corrections to the form factor
\cite{AEIP}. The polynomial part of this form factor is given by 
$W_+^{pol}=G_Fm_K^2(a_++b_+z)$, where
$z=q^2/m_K^2$ and $q$ is the momentum of the intermediate photon that
gives rise to the lepton pair. The new para\-meter $a_+$ contains in
principle also $\mathcal{O}(p^6)$ corrections and it is related with the
usual $w_+$ through \cite{AEIP}
\begin{equation}
a_+=\frac{G_8}{G_F}\Big [\frac{1}{3}-w_+\Big ].
\end{equation}
A new experimental analysis of this decay \cite{E865} measured the
parameters of the $K^+\rightarrow\pi^+e^+e^-$ form factor and found 
$a_+=-0.587\pm 0.010$. With this new number we determine $w_+$ to be
$1.086$ and $(N^r_{14}(\mu)-N^r_{15}(\mu))=-0.019\pm 0.002$ at the 
scale $m_{\rho}$. One finds that the new and the old value are almost
the same. 
\newline
\hspace*{0.5cm}
The contributions of weak loop graphs to the form factors are quite
involved. To make it more transparent how the corrections from different kinds of weak
loop graphs enter into the form factors, I present the possible kinds 
of diagrams in Figures \ref{f4} and \ref{f5} and quote the 
results separately. 
\newline
\newline\hspace*{0.5cm}
All weak tadpole diagrams can be obtained from the basic
diagram (left) in Figure \ref{f4} by appending a photon on all charged
lines and on the weak vertex. 
It turns out that only intermediate charged particles 
produce non-vanishing diagrams. The tadpole part of the form factor looks very simple and reads as
\begin{eqnarray}
\mathcal{F}_{14}^{Ll} & = & \frac{-2ieG_8}{3F}\frac{1}{d-1}
\Big \{2A(m_{\pi}^2)(2d-4)+2B(q^2,m_{\pi}^2,m_{\pi}^2)(q^2-4m_{\pi}^2)
\nonumber \\
&&+A(m_K^2)(2d-4)+B(q^2,m_K^2,m_K^2)(q^2-4m_K^2)\Big \}.
\end{eqnarray}
Diagrams which can be constructed from the second diagram (right) in Figure
\ref{f4} are referred to as diagrams of topology 1. Again, one has to
append a photon on all charged lines as well as on the strong and weak vertex. This time, only 
pairs of charged pions or charged kaons may occur in the loop. The
contribution of topology 1 to the form factors is found to be very compact,
too, and it is given by:
\begin{eqnarray}
\mathcal{F}_{15}^{Ll} & = & \frac{-ieG_8}{3F}\frac{1}{d-1}
\Big \{2(2-d)A(m_{\pi}^2)-B(q^2,m_{\pi}^2,m_{\pi}^2)(q^2-4m_{\pi}^2)
\nonumber \\
&&+(2-d)A(m_K^2)-\frac{1}{2}B(q^2,m_K^2,m_K^2)(q^2-4m_K^2)\Big \}.
\end{eqnarray}
The diagrams considered so far produce form factors that are symmetric in
the pion momenta $p_1$ and $p_2$. Besides, apart from $A$ functions only
$B(q^2,m^2,m^2)$ occurs and one can easily check that all these
contributions vanish for an on-shell photon. 
\newline\hspace*{0.5cm}
The decay amplitude 
is completed with the contributions from diagrams belonging to 
topologies 2 and 3. These diagrams are obtained from the basic graphs in 
Figure \ref{f5} through the same steps as before. The expressions that one 
obtains from these graphs are rather involved, thus I will not present the 
results explicitly in terms of the standard scalar loop functions $A$, $B$,
$C$ defined in Appendix A. 
It is the contributions of these diagrams that introduce the asymmetry in $p_1$ and $p_2$ in the $\mathcal{O}(p^4)$ 
form factors.  
The possible pairs of particles in the loop are $(\pi^0,K^-)$,
$(\eta_8,K^-)$, $(K^0_1,\pi^-)$ and $(K^0_2,\pi^-)$ for topology 2. The
particles for topology 3 are the corresponding charge conjugated ones.
It turns out that the diagrams with the internal combination
$(K^0_2,\pi^-)$ vanish. The contributions from the other possible
combinations to the form factor $\mathcal{F}^L_1$ are given in 
$\mathcal{F}_{16}^{Ll}$ in Appendix B, Eq. (\ref{FF1B}). 
\newline\hspace*{0.5cm}
By extracting only the explicit poles of the total loop contribution, one 
finds that all divergences are proportional to $q^2$, which corresponds to 
the counterterm parts of expressions (\ref{FFLl12}) and (\ref{FFLl13}). 
Furthermore, this shows that the loop amplitude of the corresponding 
radiative decay is finite \cite{ENP1,ENP2}. Finally, the complete form 
factor $\mathcal{F}^L_1$ is given by
\begin{equation}
\mathcal{F}^L_1=\mathcal{F}^{Lt}_1+\mathcal{F}^{Ll}_{11}+\mathcal{F}^{Ll}_{12}+
\mathcal{F}^{Ll}_{13}+\mathcal{F}^{Ll}_{14}+\mathcal{F}^{Ll}_{15}+
\mathcal{F}^{Ll}_{16}.
\end{equation}
As already stated, the corresponding form factor $\mathcal{F}^{L}_2$
is obtained from $\mathcal{F}^L_1$ through the substitution 
$p_1\leftrightarrow p_2$.
\subsection{\boldmath $K^+\rightarrow\pi^+\pi^0\gamma^{\star}$ Amplitudes}
The general structure of the amplitude stays the same as in (\ref{AMP}), but
there is no symmetry relation between the electric form factors
anymore. $p_1$ is now the momentum of the $\pi^+$ and $p_2$ the momentum of 
the $\pi^0$, respectively. In the limit of isospin symmetry, the octet 
tree-level amplitude vanishes, hence we relax the approximation of equal 
masses of charged and neutral pions at the tree level and take the $27$-plet 
coupling into account, too. As already anticipated in Sect. 2, throughout 
the following analysis we will use again the canonical values 
of $|G_8|$ and $G_{27}/G_8$ which are derived from the tree level.
\newline
\hspace*{0.5cm}
The tree-level form factors arise from the corresponding diagrams in Figure 
\ref{f1}, where one replaces $K_1^0$ with $K^+$ and $\pi^-$ with $\pi^0$
and puts the photon into the right places. In addition, the 
corresponding tree-level contributions with the weak coupling constant 
$G_{27}$ will be regarded, too. Terms proportional to $G_8$ are 
clearly suppressed because of approximate isospin symmetry, thus the actual 
value of $G_8$ is not of too much importance for the tree level. The 
lowest-order amplitude reads as  
\begin{eqnarray}
\label{A+tree}
\mathcal{F}^{+t}_1 & = & 2ieG_8F(m_{\pi^+}^2-m_{\pi^0}^2)\Big\{\frac{1}
{2qp_1+q^2}+\frac{1}{q^2-2qp}\Big\}\nonumber \\
&&+\frac{2ieG_{27}F}{3}(5m_{K^+}^2-7m_{\pi^+}^2+2m_{\pi^0}^2)
\frac{2qp_2}{(q^2+2qp_1)(q^2-2qp)}, \nonumber \\
\mathcal{F}^{+t}_2 & = & \frac{2ieF}{q^2-2qp}\Big[G_8(m_{\pi^+}^2-m_{\pi^0}^2)
-\frac{2G_{27}}{3}(5m_{K^+}^2-7m_{\pi^+}^2+2m_{\pi^0}^2)\Big ]. 
\end{eqnarray}
\hspace*{0.5cm}
The magnetic form factor at lowest order ($\mathcal{O}(p^4)$) is derived
from the Lagrangian (\ref{L4}) with the counterterms in (\ref{LMag}) and the 
WZW Lagrangian (\ref{LWZW}). It is necessarily finite and one calculates
\begin{equation}
\label{Amag+}
\mathcal{M}^+=\frac{eG_8}{4\pi^2F}(2-3a_2+6a_3),
\end{equation}
where the 2 comes from the Lagrangian (\ref{LWZW}). Again, the values of 
these magnetic LECs are unknown, but comparison with the corresponding
radiative $K^+$ decay \cite{BIJ,ENP2} shows that the magnetic form factor 
$\mathcal{M}^+$ also appears there. This 
suggests to use results from the E787 experiment \cite{E787} to estimate
the combination of LECs in (\ref{Amag+}). In this experiment on the
corresponding radiative $K^+$ decay, a branching ratio from direct emission 
$\mbox{BR}(K^+\rightarrow\pi^+\pi^0\gamma;\mbox{DE},55\;\mbox{MeV}
<T_{\pi^+}<90\;\mbox{MeV})=[4.7\pm 0.8\mbox{(stat.)} \pm 0.3\mbox{(syst.)}]\cdot
10^{-6}$ is reported. Under the rather reasonable assumption that direct 
emission is entirely due to the magnetic amplitude, one can extract a value 
for the whole combination of LECs in (\ref{Amag+}). Of course, this does
not take into account energy dependent corrections, but 
this is at the moment the best one can do. Moreover, the experimental data 
seem to indicate that neglect of energy dependent higher-order terms 
does not do much harm to the magnetic amplitude.
The authors also find no evidence for any electric direct emission in the
decay \cite{E787}. The combination of magnetic LECs in (\ref{Amag+}) can be 
extracted from the radiative decay ($q^2=0$) by using 
$\mathcal{A}(K^+\rightarrow \pi^+\pi^0\gamma,DE)=\mathcal{M}^+
\epsilon^{\mu\nu\rho\sigma}p_{1\nu}p_{2\rho}q_{\sigma}
\epsilon^{\star}_{\mu}(q)$: 
\begin{equation}
\label{A4est}
|2-3a_2+6a_3|=|A_4|=2.26\pm 0.25.
\end{equation}
\hspace*{0.5cm}
Turning to next-to-leading-order corrections to the electric form factors,
I start again with the discussion of contributions from strong loops and 
strong counterterm diagrams. From now on, strong isospin is conserved and 
$27$-plet corrections are neglected. 
\newline\hspace*{0.5cm}
Similar to the analysis in Sect. 3.1, diagrams with strong loops and strong 
counterterm vertices are obtained from insertions of loops and vertices in 
propagators or external lines in $K^+\rightarrow\pi^+\pi^0\gamma^{\star}$ 
tree-level diagrams derived from the graphs in Figure \ref{f1} 
by replacing $K^0_1\rightarrow K^+$ and $\pi^-\rightarrow \pi^0$. 
Wave function renormalization diagrams are not considered because the 
tree-level octet amplitude for $K^+\rightarrow\pi^+\pi^0\gamma^{\star}$ 
vanishes in the isospin limit.
The necessary insertions are obtained from the diagrams in Figure \ref{f2}, 
where $M$ denotes this time $K^+$, $\pi^+$ or $\pi^0$ and $N$ denotes
$\pi^+$, $\pi^0$, $K^+$, $K^0$, $\bar{K}^0$ or $\eta_8$,
respectively. Here, $P$ denotes $K^+$ or $\pi^+$. Introducing equivalent 
replacements as in Sect. 3.1 one obtains the strong corrections to the 
next-to-leading-order form factors given by:   
\begin{eqnarray}
\label{F+l11}
\mathcal{F}^{+l}_{11} &=& \frac{-ieG_8}{F} \Big \{-4q^2L_9-\frac{d-2}{d-1}
\Big [2A(m_{\pi}^2)+A(m_K^2)\Big ]\nonumber \\
&&+\frac{1}{d-1}\Big [(4m_{\pi}^2-q^2)B(q^2,m_{\pi}^2,m_{\pi}^2)+
\frac{1}{2}(4m_K^2-q^2)B(q^2,m_K^2,m_K^2)\Big ]\Big \},\nonumber \\
\mathcal{F}^{+l}_{21} &=& 0.
\end{eqnarray} 
The vanishing of $\mathcal{F}^{+l}_{21}$ is related to C invariance of the 
strong Lagrangian. Although counterterms proportional to $L_4$, $L_5$ and $L_9$ 
are allowed to contribute, only the $L_9$ term (\ref{CTs}) survives the 
summation of all contributions. 
The divergent part of the LEC $L_9$ removes the divergences of the result in 
(\ref{F+l11}). The form factor $\mathcal{F}^{+l}_{11}$ vanishes in the limit 
of an on-shell photon. 
\newline\hspace*{0.5cm}
Substituting the relevant weak counterterm vertex from (\ref{L4}) and 
(\ref{LEle}) for the lowest-order vertex in the direct emission diagram in 
Figure \ref{f1} and making the necessary particle replacements, one 
calculates this local contribution to the form factors:
\begin{eqnarray}
\label{FF+l12}
\mathcal{F}^{+l}_{12}&=&\frac{-ieG_8}{3F}\Big [-6qp_2(N_{14}-N_{15}-N_{16}-
N_{17})-4q^2(N_{14}-N_{15})\Big ],\nonumber \\
\mathcal{F}^{+l}_{22}&=&\frac{-ieG_8}{3F}\Big
[6qp_1(N_{14}-N_{15}-N_{16}-N_{17})-2q^2(N_{14}+2N_{15})\nonumber \\
&&+6q^2(N_{16}-N_{17})\Big ].
\end{eqnarray}
One recovers the finite combination $(N_{14}-N_{15}-N_{16}-N_{17})$ of the
corresponding radiative decay and the structure that is governed by gauge 
invariance \cite{ENP1,ENP2}. Divergences arising from weak loop diagrams are 
removed by the combinations of LECs proportional to $q^2$ in
(\ref{FF+l12}). Using the determined value of $(N^r_{16}-N_{17})$ and 
appealing to some models for weak low-energy couplings, the whole finite 
decay amplitude contains in the end only the unknown combination 
$(N^r_{14}+2N^r_{15})$ from the form factor $\mathcal{F}^{+l}_{22}$ in
(\ref{FF+l12}). I will come back to this later. At this point, it should
be mentioned that a similar combination of the same weak LECs, namely 
$(2N^r_{14}+N^r_{15})$, appears in the decay 
$K_L\rightarrow\pi^0\pi^0\gamma^{\star}$ considered in Ref. \cite{FUN}. 
\newline\hspace*{0.5cm}
Weak tadpole diagrams can be constructed from the basic
diagram in Figure \ref{f4} 
(with $K^0_2\rightarrow K^+$, $\pi^-\rightarrow \pi^0$) by following the 
same procedure as in Sect. 3.1 and one obtains
\begin{eqnarray}
\label{FF+l13}
\mathcal{F}^{+l}_{13}&=&\frac{-ieG_8}{3F}\frac{1}{d-1}\Big \{(2d-4)
[A(m_{\pi}^2)+A(m_K^2)]\nonumber \\
&&+B(q^2,m_{\pi}^2,m_{\pi}^2)(q^2-4m_{\pi}^2)+B(q^2,m_K^2,m_K^2)
(q^2-4m_K^2)\Big \},\nonumber \\
\mathcal{F}^{+l}_{23}&=&\frac{-ieG_8}{3F}\frac{1}{d-1}\Big \{(2d-4)
[A(m_{\pi}^2)-A(m_K^2)]\nonumber \\
&&+B(q^2,m_{\pi}^2,m_{\pi}^2)(q^2-4m_{\pi}^2)-B(q^2,m_K^2,m_K^2)
(q^2-4m_K^2)\Big \}.
\end{eqnarray}   
One finds that $\eta_8$ loops do not contribute. In case of an on-shell photon, 
expressions (\ref{FF+l13}) vanish. 
\newline\hspace*{0.5cm}
Diagrams of the topology 1 are constructed from the right diagram in
Figure \ref{f4} (with $K^0_2\rightarrow K^+$, $\pi^-\rightarrow \pi^0$) 
replacing the charged meson pairs in the loop with $(\pi^+,\pi^0)$, 
$(K^+,\bar{K}^0)$, or with $(\pi^+,\eta_8)$ (in an appropriate momentum 
convention). The last combination of intermediate particles vanishes in the 
isospin limit. Appending a photon where it is 
possible and summing up the diagrams, one finds a compact result involving 
only $A$ and $B(q^2,m^2,m^2)$:
\begin{eqnarray}
\label{FF+l14}
\mathcal{F}^{+l}_{14}&=&\frac{-ieG_8}{F(d-1)}
\Big \{(d-2)\Big [A(m_K^2)+\frac{4}{3}A(m_{\pi}^2)\Big ]\nonumber \\
&&+\frac{1}{2}B(q^2,m_K^2,m_K^2)(q^2-4m_K^2)+\frac{2}{3}B(q^2,m_{\pi}^2,
m_{\pi}^2)(q^2-4m_{\pi}^2)\Big \},\nonumber \\
\mathcal{F}^{+l}_{24}&=&\frac{-ieG_8}{F(d-1)}\Big \{(2-d)
\Big[A(m_K^2)+\frac{2}{3}A(m_{\pi}^2)\Big ]\nonumber \\
&&+\frac{1}{2}B(q^2,m_K^2,m_K^2)(4m_K^2-q^2)+\frac{1}{3}
B(q^2,m_{\pi}^2,m_{\pi}^2)(4m_{\pi}^2-q^2)\Big \}.
\end{eqnarray}
Expressions (\ref{FF+l14}) vanish in the on-shell limit and their divergences are
clearly proportional to $q^2$. The last and by far most voluminous
contributions to the electric $\mathcal{O}(p^4)$ form factors come from
diagrams of the topologies 2 and 3 which can be derived from the basic
diagrams in Figure \ref{f5} as before. Possible virtual pairs 
are $(\pi^0,K^-)$, $(\eta_8,K^-)$, $(K^0,\pi^-)$ and $(K^0,\pi^0)$, 
$(K^0,\eta_8)$, $(K^+,\pi^+)$, respectively. The obtained results, 
labelled as $\mathcal{F}^{+l}_{15,25}$ and
$\mathcal{F}^{+l}_{16,26}$, are listed in Appendix C, (\ref{FF+l15}), 
(\ref{FF+l25}) and (\ref{FF+l16}), (\ref{FF+l26}). The complete form factor 
$\mathcal{F}^+_1$ is finally given by 
\begin{equation}
\mathcal{F}^+_1=\mathcal{F}^{+t}_1+\mathcal{F}^{+l}_{11}+\mathcal{F}^{+l}_{12}+
\mathcal{F}^{+l}_{13}+\mathcal{F}^{+l}_{14}+\mathcal{F}^{+l}_{15}+
\mathcal{F}^{+l}_{16}.
\end{equation}
$\mathcal{F}^+_2$ is obtained from the corresponding sum.
\section{Numerical analysis} 
\subsection{Decay width}
The decay width for the processes in question is given by the following
standard formula:
\begin{eqnarray}
\Gamma(K\rightarrow\pi_1\pi_2e^+e^-)&=&\frac{m_e^2}{128\pi^8m_K}
\int\frac{d^3p_1}{2E_1}\frac{d^3p_2}{2E_2}\frac{d^3k_+}{2E_+}\frac{d^3k_-}
{2E_-}\delta^{(4)}(p_f-p_i)\sum_{spins}|\mathcal{A}|^2,\nonumber \\ 
&&
\end{eqnarray}
where $p_1$ is always the momentum of the positive pion and $p_2$ refers to
the corresponding other pion, $\pi^-$ or $\pi^0$. As usual, $p_{i,f}$ denote
the sums of ingoing and outgoing momenta, respectively. In fact, 
$p^{\mu}_i=(m_K,0,0,0)$. The squared transition amplitude for both decays
in question reads as follows:
\begin{eqnarray}
\label{Asqr}
\sum_{spins}|\mathcal{A}|^2&=&\frac{e^2}{m_e^2q^4}\bigg \{-(m_e^2+k_+k_-)
\Big [|\mathcal{F}_1|^2p_1^2+|\mathcal{F}_2|^2p_2^2+p_1p_2
(\mathcal{F}_1\mathcal{F}_2^{\star}+\mathcal{F}_1^
{\star}\mathcal{F}_2)\Big ]\nonumber \\
&&+2|\mathcal{F}_1|^2k_+p_1k_-p_1+2|\mathcal{F}_2|^
2k_+p_2k_-p_2+(\mathcal{F}_1\mathcal{F}_2^{\star}+\mathcal{F}_1^{\star}
\mathcal{F}_2)\nonumber \\
&&(k_+p_1k_-p_2+k_-p_1k_+p_2)\bigg \}+\frac{e^2|\mathcal{M}|^2}{m_e^2q^4}
\bigg \{(-m_e^2+k_+k_-)\Big [p_1^2qp_2^2\nonumber \\
&&+p_2^2qp_1^2+q^2p_1p_2^2-p_1^2p_2^2q^2-2p_1p_2qp_1qp_2\Big
]+2qk_-qk_+\nonumber \\
&&(p_1^2p_2^2-p_1p_2^2)+2k_-p_1k_+p_1(p_2^2q^2-qp_2^2)+2k_-p_1k_+p_2
(p_1^2q^2-qp_1^2)\nonumber \\
& &+2(qp_1qp_2-q^2p_1p_2)(k_+p_1k_-p_2+k_+p_2k_-p_1)+
2(p_1p_2qp_2-p_2^2qp_1)\nonumber \\
& &(k_+p_1k_-q+k_-p_1k_+q)+
2(p_1p_2qp_1-p_1^2qp_2)(k_+p_2k_-q+k_-p_2k_+q)\bigg \}\nonumber \\
& &+\frac{e^2}{m_e^2q^4}\epsilon^{\mu\nu\rho\sigma}
k_{-\mu}p_{1\nu}p_{2\rho}k_{+\sigma}\bigg
\{(k_+p_1-k_-p_1)(\mathcal{F}_1^{\star}\mathcal{M}+\mathcal{F}_1\mathcal{M}^
{\star})\nonumber \\
& &+(k_+p_2-k_-p_2)(\mathcal{F}_2^{\star}\mathcal{M}+\mathcal{F}_2
\mathcal{M}^{\star})\bigg \}.
\end{eqnarray}
The structure of (\ref{Asqr}) implies that there is no interference between
electric and magnetic form factors in the decay widths of these decays. 
Additionally, one finds for the decay of the $K_L$ that there is no 
interference between electric form factors of lowest and next-to-leading order,
too. This feature is due to their different behaviour under exchange of
pion momenta. $K_L$ branching ratios thus consist of three distinct 
contributions. The more general case of interference between electric 
form factors of different orders is present in the $K^+$ decay.
Phase space integrations are performed numerically with the Fortran event
generator RAMBO \cite{EKS}.
\subsection{Numerical analysis of \boldmath $K_L\rightarrow \pi^+\pi^-e^+e^-$}
\begin{figure}[h]
\vspace*{-18.5cm}
\hspace*{2.0cm}
\resizebox{1.2\textwidth}{!}{
\includegraphics{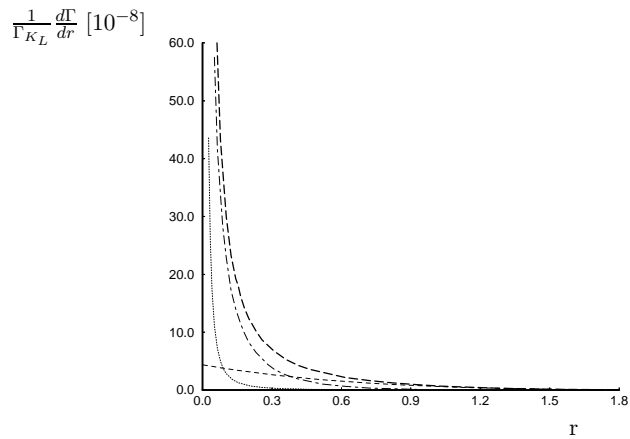}}
\caption{Differential decay width $\frac{1}{\Gamma_{K_L}}\frac{d\Gamma}
{dr}$ for $(N^r_{16}(m_{\rho})-N_{17})=-0.014$; $r:=q^2/m_{\pi}^2$. The dotted
line is the tree-level contribution, the dashed line refers to loops and
counterterms, the dot-dashed line refers to the magnetic part. The thick 
long-dashed line is the sum. $\Gamma_{K_L}$ is the total width of the $K_L$. For cuts
with $q^2>(130\;\mbox{MeV})^2$ the differential decay width is dominated by
the contributions of the electric $\mathcal{O}(p^4)$ amplitude, i.e. 
loops and electric counterterms.}
\label{f6}
\end{figure}

In the following, branching ratios (BRs) for different cuts in $q^2$, 
i.e. for different lower bounds on $(k_+ +k_-)^2$, and the BR over the entire
phase space are listed. $q^2$ may vary bet\-ween $4m_e^2$ and 
$(m_K-2m_{\pi})^2$. Throughout this analysis, the central values of 
experimental numbers are used for the branching ratios with certain cuts 
in $q^2$. The error of a branching ratio is only given if it is calculated 
over the whole phase space. It should be pointed out 
(compared to Ref. \cite{SAV}) that the KTeV data, on which I will rely in
the following analysis, are corrected for the entire phase space 
\cite{KTEV2,BAR,KETT}. 
\begin{table}[h]
\caption{Magnetic and tree-level contributions to the branching ratio of 
$K_L\rightarrow \pi^+\pi^-e^+e^-$ for different cuts in $q^2$ and for the 
entire phase space.}
\vspace*{0.2cm}
\hspace*{1.9cm}
\begin{tabular}{lll}
\hline\hline\noalign{\smallskip}
$q^2 > (\mbox{MeV}^2)$ & Magnetic BR $[10^{-8}]$ & Tree-level BR 
$[10^{-8}]$ \\ 
\noalign{\smallskip}\hline\noalign{\smallskip}
$2^2$ & $18.20$ & $9.8$\\ 
$10^2$ & $9.31$ & $2.95$\\ 
$20^2$ & $5.61$ & $1.33$\\ 
$30^2$ & $3.65$ & $0.71$\\ 
$40^2$ & $2.42$ & $0.41$\\ 
$60^2$ & $1.06$ & $0.16$\\ 
$80^2$ & $0.44$ & $0.061$\\ 
$100^2$ & $0.16$ & $0.024$\\ 
$120^2$ & $0.053$ & $0.009$\\ 
$180^2$ & $0.00025$ & $0.0001$\\ 
$\mbox{entire p.s.}$ & $21.2 \pm 9.0$ & $12.8 \pm 1.0$\\ 
\noalign{\smallskip}\hline
\end{tabular}
\label{TLmagtree}
\end{table}
\newline
\hspace*{0.5cm}
Using ansatz (\ref{AmagKTeV}) and the experimental numbers of \cite{KTEV2} as input for
the magnetic contribution to the branching ratio, I find the results 
collected in Table \ref{TLmagtree}. Neglecting the energy dependent part in 
(\ref{AmagKTeV}) one reproduces the results in Ref. \cite{ESW}. Obviously, 
consideration of the energy dependent magnetic form factor in 
(\ref{AmagKTeV}) increases the results compared to a constant magnetic form
factor, particularly for low cuts \cite{ESW}. Unfortunately, the errors of the
parameters entering into the magnetic contribution to the BR are rather 
large \cite{KTEV2}, thus the magnetic branching ratio over the whole phase 
space has a considerable uncertainty.
\newline
\hspace*{0.5cm}
The tree-level form factors of (\ref{ALtree}) give rise to the results
collected in column three of Table \ref{TLmagtree}. Comparison with the 
results in \cite{ESW} shows that the obtained numbers are rather 
different, but this is due to different values of $F$ and $G_8$. Here, we use
$F=92.4\;\mbox{MeV}$ and the canonical $|G_8|$.
The error of the BR over the entire phase space in the last line comes from 
numerics and reflects the $1/q^4$ behaviour of the squared amplitude.
Table \ref{TLmagtree} also shows very clearly the importance of the $q^2$ 
range between $4m_e^2$ and $4\;\mbox{MeV}^2$ that was not considered in 
\cite{SAV}. The importance of this small $q^2$ range is understood from the 
plot of the differential decay widths of the individual contributions to
the decay in Figure \ref{f6}.
\newline
\hspace*{0.5cm}
The $\mathcal{O}(p^4)$ electric form factors depend via (\ref{FFLl13}) 
on $(N^r_{14}(\mu)-N^r_{15}(\mu)-3(N^r_{16}(\mu)-N_{17}))=:X(\mu)$, hence  
electric next-to-leading-order contributions to the branching ratios 
are given as functions of $X$. The derived branching ratios are listed 
in Table \ref{TL4X} and allow in principle for an extraction of the whole
combination of LECs. $X$ is counted in units of $10^{-2}$. 
\begin{table}[h]
\caption{Contributions of loops and electric counterterms to the branching 
ratio of $K_L\rightarrow\pi^+\pi^-e^+e^-$ for different cuts in $q^2$ and
for the entire phase space, given as functions of the combination  
$(N^r_{14}(\mu)-N^r_{15}(\mu)-3(N^r_{16}(\mu)-N_{17}))=:X(\mu)[10^{-2}]$ 
of weak LECs. The error is due to the uncertainty of $L^r_9$.}
\vspace*{0.2cm}
\hspace*{2.25cm}
\begin{tabular}{ll}
\hline\hline\noalign{\smallskip}
$q^2 > (\mbox{MeV}^2)$ & Loops$+$Counterterms BR $[10^{-8}]$ \\ 
\noalign{\smallskip}\hline\noalign{\smallskip}
$2^2$ & $0.87+0.46X+0.06X^2$\\ 
$10^2$ & $0.86+0.45X+0.06X^2$\\ 
$20^2$ & $0.84+0.44X+0.06X^2$\\ 
$30^2$ & $0.80+0.42X+0.05X^2$\\ 
$40^2$ & $0.75+0.39X+0.05X^2$\\ 
$60^2$ & $0.61+0.32X+0.04X^2$\\ 
$80^2$ & $0.46+0.23X+0.03X^2$\\ 
$100^2$ & $0.30+0.16X+0.02X^2$\\ 
$120^2$ & $0.18+0.09X+0.01X^2$\\ 
$180^2$ & $0.007+0.003X+0.0004X^2$\\ 
$\mbox{entire p.s.}$ & $0.87\pm 0.19+(0.46\pm 0.03)X+0.06X^2$\\
\noalign{\smallskip}\hline
\end{tabular}
\label{TL4X}
\end{table}
The results collected in Table \ref{TL4X} exhibit very clearly the 
\newpage\noindent
entire contribution of the involved weak local counterterms to the electric order 
$p^4$ branching ratios. 
\newline\hspace*{0.5cm}
With the help of Eq. (\ref{W+}), we extract for the combination 
$(N^r_{14}(m_{\rho})-N^r_{15}(m_{\rho}))$ a central value of $-0.019$; 
therefore, the next-to-leading-order electric form factors effectively
depend only on $(N^r_{16}(\mu)-N_{17})=:x(\mu)$ and electric 
contributions of order $p^4$ to the branching ratios can also be expressed
as functions of $x$, again counted in units of $10^{-2}$. The derived 
branching ratios are listed in Table \ref{TL4x}. The error in Table
\ref{TL4x} was estimated by taking into account the uncertainties of
$L^r_9$ and $(N^r_{14}-N^r_{15})$.
\begin{table}[h]
\caption{Contributions of loops and electric counterterms to the branching 
ratio of $K_L\rightarrow\pi^+\pi^-e^+e^-$ for different cuts in $q^2$ and 
for the entire phase space, given as functions of the combination 
$(N_{16}^r-N_{17})=:x\;[10^{-2}]$ of weak LECs. $(N^r_{14}(m_{\rho})-N^r_{15}
(m_{\rho}))=-0.019$ was used.}
\vspace*{0.2cm}
\hspace*{2.4cm}
\begin{tabular}{ll}
\hline\hline\noalign{\smallskip}
$q^2 > (\mbox{MeV}^2)$ & Loops$+$Counterterms BR $[10^{-8}]$ \\ 
\noalign{\smallskip}\hline\noalign{\smallskip}
$2^2$ & $0.22-0.68x+0.54x^2$\\ 
$10^2$ & $0.22-0.67x+0.53x^2$\\ 
$20^2$ & $0.21-0.66x+0.52x^2$\\ 
$30^2$ & $0.20-0.63x+0.49x^2$\\ 
$40^2$ & $0.19-0.59x+0.46x^2$\\ 
$60^2$ & $0.16-0.48x+0.37x^2$\\ 
$80^2$ & $0.12-0.36x+0.27x^2$\\ 
$100^2$ & $0.08-0.24x+0.18x^2$\\ 
$120^2$ & $0.05-0.14x+0.10x^2$\\ 
$180^2$ & $0.002-0.005x+0.004x^2$\\ 
$\mbox{entire p.s.}$ & $0.22\pm 0.11-(0.68\pm 0.16)x+0.54x^2$\\ 
\noalign{\smallskip}\hline
\end{tabular}
\label{TL4x}
\end{table}
\newline\hspace*{0.5cm}
The numbers in Tables \ref{TL4X} and \ref{TL4x} cannot be compared immediately 
to the results in \cite{ESW}, since the corresponding branching ratios 
were expressed as functions of a different combination of LECs, $w_L$ 
\cite{ESW}. It turns out that $w_L$ is related to the used 
$N_i$ through $w_L=8\pi^2[-N_{14}+N_{15}+N_{16}-N_{17}]$. 
\newline\hspace*{0.5cm}
It is obvious from Tables \ref{TL4X} and \ref{TL4x} that the electric 
$\mathcal{O}(p^4)$ contributions are nearly insensitive to changes of the 
cut below $\sim(40\;\mbox{MeV})^2$. This feature becomes also clear from 
inspection of Figure \ref{f6}.  
\newline\hspace*{0.5cm}
Theory finally predicts as central value of 
$BR(K_L\rightarrow\pi^+\pi^-e^+e^-)$ over the entire phase 
space $[21.2+12.8+0.87+0.46X+0.06X^2]\cdot 10^{-8}$, with
$X=(N^r_{14}-N^r_{15}-3(N^r_{16}-N_{17}))[10^{-2}]$ . Comparison with the 
branching ratio obtained in Ref. \cite{SW},
$\mbox{BR}=[18\;\mbox{(magn.)}+13\;\mbox{(tree)}+0.4\;\mbox{(CR)}]
\cdot 10^{-8}$, shows that inclusion of the magnetic form factor of Ref. 
\cite{KTEV2} increases the magnetic BR considerably. Also the total 
$\mathcal{O}(p^4)$ electric contribution of Table \ref{TL4X} changes the 
result to some extent.
\newline
\hspace*{0.5cm}
In the following, the obtained theoretical BR over the entire phase space 
from Table \ref{TL4x} will be compared to the most recent available data to 
extract values for $(N^r_{16}(m_{\rho})-N_{17})$. It should be mentioned 
that possible values of the related combination $w_L$ of LECs were estimated
in Ref. \cite{SAV} by comparing with the then recent BR. However, a 
theoretical cut of $q^2=(2\;\mbox{MeV})^2$ was applied and the energy 
dependence of the magnetic form factor (\ref{AmagKTeV}) was not taken into 
account. 
\newline\hspace*{0.5cm}
I focus on the data of the KTeV collaboration, but for completeness one
should mention that a Japanese group obtained a BR of 
$[4.4 \pm 1.3\mbox{(stat.)}\pm 0.5\mbox{(syst.)}]\cdot 10^{-7}$ \cite{TAK},
based on 13 events, 
and that the NA48 experiment at CERN
recently reported a preliminary BR of $(3.1 \pm 0.3)\cdot 10^{-7}$
\cite{MAZ,BIZZ}. In the last years, the KTeV result for the branching ratio 
was subject to numerous analyses and the errors improved quite a lot. 
The first published BR was based on a sample
of 46 events and it was found to be $[3.2\pm 0.6\mbox{(stat.)}\pm 0.4
\mbox{(syst.)}]\cdot 10^{-7}$ \cite{KTEV1}. A new analy\-sis based on the
full 1997 data set reported a BR of $[3.32 \pm 0.14\mbox{(stat.)}\pm 0.28 
\mbox{(syst.)}]\cdot 10^{-7}$ with a much better statistical error
\cite{KTEV3}. I am going to use the latest available (preliminary) numbers 
which were again obtained from the 1997 data set by considering the 
parametrization in (\ref{AmagKTeV}) \cite{BAR,KETT}: $\mbox{BR}=[3.63\pm 0.11\mbox{(stat.)}
\pm 0.14\mbox{(syst.)}]\cdot 10^{-7}$.  
\newline\hspace*{0.5cm}
It is clear that it is not possible to determine unambiguously the value 
of $x$ only by comparison with the experiment. The two possible values of 
$x$ are
\begin{eqnarray}
\label{Result}
(N^r_{16}(m_{\rho})-N_{17})_1&=\;x_1\;=&(2.7\pm 3.6)\cdot 10^{-2},\nonumber \\
(N^r_{16}(m_{\rho})-N_{17})_2&=\;x_2\;=&(-1.4\pm 3.6)\cdot 10^{-2}.
\end{eqnarray}
The large error is mostly ($\sim 80\%$) due to the uncertainty of the 
magnetic BR. There is even a small overlap of the two ranges of $x_1$ and 
$x_2$. Moreover, it should be stressed that comparison with the BR over the entire phase 
space is not the best possibi\-lity to extract values for the LECs, since
the BR over the whole phase space is dominated by the tree level and the 
magnetic amplitude. In addition, Figure \ref{f6} suggests that one could 
extract a value to a better precision for much higher cuts in $q^2$, but
this is not possible at the moment. On the other hand, such an extraction 
would suffer from smaller 
statistics. Nevertheless, the central values in (\ref{Result}) are very 
different and one can appeal to models of weak counter\-term couplings to 
distinguish between the two solutions. The Weak Deformation Model (WDM) and 
the Factorization Model (FM) \cite{EKW} make predictions about the involved
LECs, but apart from a free parameter of the FM, $N^r_{14}$ and $N^r_{16}$ 
depend in both models on the contact term coupling $H_{1}$ from
the strong counterterm Lagrangian \cite{GL2}. Therefore, it is necessary to 
compare $(N^r_{14}-N^r_{15})-(N^r_{16}-N_{17})$ to the experimental values,
since in this combination $H_{1}$ drops out.
\begin{table}[h]
\caption{Comparison of model predictions (WDM, FM)
and the two possible values extracted from data for the combination 
$(N^r_{14}-N^r_{15})-(N^r_{16}-N_{17})$ at $m_{\rho}$.}
\vspace*{0.2cm}
\hspace*{2.0cm}
\begin{tabular}{llll}
\hline\hline\noalign{\smallskip}\noalign{\smallskip}
Model & Pred. & $(N^r_{14}-N^r_{15})-x_1$ & 
$(N^r_{14}-N^r_{15})-x_2$\\ 
\noalign{\smallskip}\hline
WDM & $-0.004$ & & \\ 
FM & $-0.007k_f$ & \raisebox{1.0ex}[-1.0ex]{$-0.046\pm 0.036$} &
\raisebox{1.0ex}[-1.0ex]{$-0.005\pm 0.036$} \\ 
\noalign{\smallskip}\hline
\end{tabular}
\label{Models}
\end{table}
The comparison of prediction and experiment is given in Table \ref{Models}.
$k_f$ parametrizes the factorization hypothesis and is expected to be of 
$\mathcal{O}(1)$. Comparison with the results in Table \ref{Models} gives 
$k_{f1}\simeq 6.4\pm 5.0$ and $k_{f2}\simeq 0.7\pm 5.0$, respectively.
First of all, it is remarkable to find the central value of $x_2$ to be in
such good agreement with the predictions of the two models. Secondly, the
errors are big enough to dampen too much enthusiasm, but in any case the
solution $(N^r_{16}(m_{\rho})-N_{17})=-0.014$ is clearly favoured by both
models. Moreover, both models and also the results of \cite{BP} suggest 
that $N_{17}$ vanishes individually, so one can even go 
one step further and assume that
\begin{equation}
\label{Assump}
N^r_{16}(m_{\rho})=(-1.4\pm 3.6)\cdot 10^{-2}\;,\;\;N_{17}=0.
\end{equation}
Additional support comes from the null measurement of interference between 
electric direct emission and brems\-strahlung in the $K^+\rightarrow 
\pi^+\pi^0\gamma$ amplitude in Ref. \cite{E787}, since this
indicates that the combination $(N^r_{14}-N^r_{15}-N^r_{16}-N_{17})\simeq 0$  
or very small \cite{ENP1,ENP2}.
Assumption (\ref{Assump}), however, is only true for a 
certain class of models and one should take into account other approaches, 
too, e.g. the modified FM (FMV) approach \cite{DAP1,DAP} which was
originally introduced to estimate $\mathcal{O}(p^6)$ corrections to
radiative kaon decays. This model can also be used to parametrize LECs and 
it predicts in general a $N_{17}$ different from zero. In any case,
solution $x_2$ is also supported  by the results in \cite{DAP}. We
will consider the FMV more closely in the next section.
\newline\hspace*{0.4cm}
If one's trust in the models used above were big enough, one could even use 
(\ref{Assump}) to calculate the contact term coupling $H_{1}$ of Ref. 
\cite{GL2} and, as a consequence, calculate $N^r_{14}$ and $N^r_{15}$, but this does
not seem to make much sense. In any case, assumption (\ref{Assump}) serves
as a solid starting point for the analysis of the $K^+$ decay. 
\subsection{Numerical analysis of \boldmath $K^+\rightarrow \pi^+\pi^0e^+e^-$}
As in the previous case, there is no interference between magnetic
and electric parts of the amplitude. Because of the absence of a symmetry
relation between the electric form factors as in the $K_L$ decay, however, this
time there is interference between the tree-level amplitude and loops and 
electric counterterms. 
\newline
\hspace*{0.5cm}
For the purely magnetic part of the branching ratio, the value of 
(\ref{A4est}) is used. The results are collected in the second column of 
Table \ref{T+magtree}. As in Sect. 4.2, I only quote the error associated 
with $|A_4|$ for the branching ratio over the entire phase space.
\begin{table}[h]
\caption{Magnetic and tree-level contributions to the 
branching ratio of $K^+\rightarrow\pi^+\pi^0e^+e^-$ for different cuts 
in $q^2$ and for the entire phase space. The error of the magnetic part 
is due to experimental uncertainties of $|A_4|$, whereas the error of the
tree-level result comes from numerics; additionally, there is an intrinsic
uncertainty because of the $G_{27}$ coupling.}
\vspace*{0.2cm}
\hspace*{1.9cm}
\begin{tabular}{lll}
\hline\hline\noalign{\smallskip}
$q^2 > (\mbox{MeV}^2)$ & Magnetic BR $[10^{-8}]$ & Tree-level BR 
$[10^{-8}]$\\ 
\noalign{\smallskip}\hline\noalign{\smallskip}
$2^2$ & $5.33$ & $254.20$\\ 
$10^2$ & $2.84$ & $74.33$\\ 
$20^2$ & $1.80$ & $32.51$\\ 
$30^2$ & $1.23$ & $17.35$\\ 
$40^2$ & $0.86$ & $10.04$\\ 
$60^2$ & $0.42$ & $3.75$\\ 
$80^2$ & $0.19$ & $1.46$\\ 
$100^2$ & $0.083$ & $0.56$\\ 
$120^2$ & $0.031$ & $0.20$\\ 
$180^2$ & $0.0002$ & $0.002$\\ 
$\mbox{entire p.s.}$ & $6.14 \pm 1.3$ & $330 \pm 15 $ \\ 
\noalign{\smallskip}\hline
\end{tabular}
\label{T+magtree}
\end{table}
\newline\hspace*{0.5cm}
In the following, I present the individual BRs due to electric form factors
of lowest and next-to-leading order as well as the total electric branching 
ratios, which allows for an extraction of the interference contribution. 
The BR due to the lowest order is generated by the tree-level form factors given 
in (\ref{A+tree}). They produce a branching ratio that is much larger than
that of $K_L\rightarrow\pi^+\pi^-e^+e^-$; it is given in the third column
of Table \ref{T+magtree}. As already mentioned, the tree-level value of $G_{27}$
is used for the numerical analysis and this clearly introduces an intrinsic 
uncertainty in the predictions. The error associated with the choice of $G_8$ is very small for the tree 
level. For loops and counter\-term contributions we choose again the canonical
$|G_8|$. For completeness, I quote
the error arising from numerics because of the $1/q^4$ behaviour of the 
tree-level amplitude. 
\newline
\hspace*{0.5cm}
The $\mathcal{O}(p^4)$ form factors contain the combinations of LECs
given in (\ref{FF+l12}). Here, I use the assumption of (\ref{Assump}) and 
thus express the derived branching ratios as functions of 
$(N^r_{14}(\mu)+2N^r_{15}(\mu))=:z(\mu)$ (in units of $10^{-2}$). 
The results are given in Table \ref{T+L4}. The total electric
contributions form the tree level, loops and counterterms are collected in 
Table \ref{T+L24}. 
\begin{table}[h]
\caption{Electric $\mathcal{O}(p^4)$ contribution to the branching ratio 
of $K^+\rightarrow\pi^+\pi^0e^+e^-$ for cuts in $q^2$ and for the entire 
phase space considering the error of (\ref{Assump}). 
$(N^r_{14}+2N^r_{15}):=z\;[10^{-2}]$.}
\vspace*{0.2cm}
\hspace*{2.5cm}
\begin{tabular}{ll}
\hline\hline\noalign{\smallskip}
$q^2 > (\mbox{MeV}^2)$ & Loops+Counterterms BR $[10^{-9}]$ \\ 
\noalign{\smallskip}\hline\noalign{\smallskip}
$2^2$ & $46+0.51z+0.042z^2$ \\ 
$10^2$ & $2.52+0.48z+0.042z^2$ \\ 
$20^2$ & $1.68+0.46z+0.041z^2$ \\
$30^2$ & $1.43+0.43z+0.039z^2$ \\ 
$40^2$ & $1.25+0.40z+0.036z^2$ \\ 
$60^2$ & $0.94+0.32z+0.030z^2$ \\
$80^2$ & $0.67+0.24z+0.022z^2$ \\ 
$100^2$ & $0.43+0.16z+0.015z^2$ \\ 
$120^2$ & $0.25+0.09z+0.009z^2$ \\ 
$180^2$ & $0.01+0.004z+0.0004z^2$ \\ 
$\mbox{entire p.s.}$ & $122\pm 134+(0.56\pm 0.27)z+0.043z^2$ \\ 
\noalign{\smallskip}\hline
\end{tabular}
\label{T+L4}
\end{table}
From this analysis it is clear that it will be very difficult to isolate 
the electric $\mathcal{O}(p^4)$ corrections to branching ratios with small
or no cuts in $q^2$. 
Again, the importance of the last step from a cut of $4\;\mbox{MeV}^2$ to 
no cut at all should be mentioned. Comparison with the previous $K_L$ 
decay shows that the tree-level contribution, although it is suppressed by 
isospin symmetry, do\-minates the BR and that it is much more important than
for the $K_L$ decay, where the tree level was $\epsilon$-suppressed. 
\newline
\hspace*{0.3cm}
Finally, using only the central values of input quantities and applying 
assumption (\ref{Assump}), $N^r_{16}(m_{\rho})=-0.014$ and $N_{17}=0$, the 
central value of the branching ratio for $K^+\rightarrow \pi^+\pi^0e^+e^-$ 
over the entire phase space is predicted to be 
$[6\;+\;378+0.27z+0.004z^2]\cdot 10^{-8}$. 
\begin{table}[h]
\caption{Contributions of the total electric part of the amplitude, 
$\mathcal{O}(p^2)$ and $\mathcal{O}(p^4)$, to the branching ratio of 
$K^+\rightarrow\pi^+\pi^0e^+e^-$ for cuts in $q^2$ and for the entire phase 
space with consideration of the error of (\ref{Assump}). 
$(N^r_{14}+2N^r_{15}):=z\;[10^{-2}]$.}
\vspace*{0.2cm}
\hspace*{2.35cm}
\begin{tabular}{ll}
\hline\hline\noalign{\smallskip}
$q^2 > (\mbox{MeV}^2)$ & Electric $\mathcal{O}(p^2)+\mathcal{O}(p^4)$ 
BR $[10^{-9}]$ \\ 
\noalign{\smallskip}\hline\noalign{\smallskip}
$2^2$ & $2745+2.69z+0.042z^2 $ \\ 
$10^2$ & $785.6+2.57z+0.042z^2 $ \\ 
$20^2$ & $354.3+2.29z+0.041z^2 $ \\
$30^2$ & $193.8+2.09z+0.039z^2 $ \\ 
$40^2$ & $115.6+1.83z+0.036z^2 $ \\ 
$60^2$ & $46.3+1.32z+0.030z^2$ \\
$80^2$ & $19.8+0.88z+0.022z^2$ \\ 
$100^2$ & $8.40+0.52z+0.015z^2$ \\ 
$120^2$ & $3.34+0.27z+0.009z^2$ \\ 
$180^2$ & $0.06+0.008z+0.0004z^2$ \\ 
$\mbox{entire p.s.}$ & $3783\pm 350+(2.74\pm 0.82)z+0.043z^2$ \\ 
\noalign{\smallskip}\hline
\end{tabular}
\label{T+L24}
\end{table}
\subsection{Dependence on counterterm models}
In Sect. 4.2 it was claimed that also the modified Factorization Model
(FMV) of Refs. \cite{DAP1,DAP} favours $(N^r_{16}(m_{\rho})-N_{17})=-0.014$. 
We will now clarify why this is so. In Refs. \cite{DAP1,DAP}, the authors 
introduced a different approach compared to the one used in \cite{EKW} to 
incorporate interactions between pseudoscalars and vector and axial-vector 
resonances. Factorization, however, was still an important ingredient. 
Originally used to estimate order $p^6$ corrections to kaon
decays, their framework was also extended to parametrize combinations of
weak LECs in terms of two positive $\mathcal{O}(1)$ parameters, $\eta_V$ and
$\eta_A$. According to \cite{DAP}, however, one finds that the weak LECs we
are interested in, $N^r_{14}$, $N^r_{15}$, $N^r_{16}$ and $N_{17}$, do not
depend on the factorization hypothesis. Hence, one should consider the
parametrizations of combinations of these LECs in terms of $\eta_V$ and
$\eta_A$ as model independent. Of course, these parametrizations of 
combinations of LECs still depend on the formalism applied to incorporate 
vector and axial-vector resonances \cite{DAP}
\footnote{I thank J. Portol\'{e}s for useful comments on this topic.}.  
\newline\hspace*{0.35cm}
In the FMV, the relations $(N^r_{14}-N^r_{15})=-0.020\eta_V+0.004\eta_A$ and 
$(N^r_{14}-N^r_{15})-3(N^r_{16}-N_{17})=-0.004\eta_V+
0.018\eta_A$ hold. Comparison with the two possible values of 
$(N^r_{16}-N_{17})$ in (\ref{Result}) gives the following results for the two 
parameters: for $(N^r_{16}(m_{\rho})-N_{17})=0.027$, one finds  
\begin{equation} 
\label{Set1}
\eta_{V_1}=-0.2\pm 1.1\;\;\;\mbox{and} \;\;\;\eta_{A_1}=-5.6\pm 5.8.
\end{equation}
Using $(N^r_{16}(m_{\rho})-N_{17})=-0.014$, one calculates for the FMV 
parameters
\begin{equation} 
\label{Set2}
\eta_{V_2}=1.3\pm 1.1\;\;\;\mbox{and} \;\;\;\eta_{A_2}=1.6\pm 5.8.
\end{equation}
Of course, the errors are very big, but even then the second pair of
values in (\ref{Set2}) fits much better than the values of (\ref{Set1}). 
Using the parametrization $(N^r_{14}-N^r_{15})-3(N^r_{16}+
N_{17})=0.05\eta_V-0.04\eta_A$, one determines 
$N_{17}=-0.009\eta_V+0.0097\eta_A=0.4\cdot 10^{-2}$. Thus, we rather find
for $N^r_{16}$ and $N_{17}$ with the values of (\ref{Set2})
\begin{eqnarray}
\label{Result1}
N^r_{16}(m_{\rho})&=&(-1.0\pm 4.6)\cdot10^{-2},\nonumber \\
N_{17}&=&(0.4\pm 4.6)\cdot10^{-2}
\end{eqnarray}
In fact, this result is not too different from the
hypothesis in (\ref{Assump}). 
\newline
\hspace*{0.5cm}
Despite the big uncertainties of the values of the parameters 
$\eta_V$ and $\eta_A$, one nevertheless can use the central values to 
calculate other LECs,
especially $N^r_{14}$ and $N^r_{15}$. Probably this is equally daring as
the option of calculating $H_{1}$, but if one truly 'believed' in the FMV 
with the results of (\ref{Set2}), one could make a parameter-free prediction 
for the electric part of the branching ratio. According to \cite{DAP}, 
we have $2N^r_{14}+N^r_{15}=0.08\eta_V$. Using the experimental
result $(N^r_{14}-N^r_{15})=-0.019$, one calculates:
\begin{eqnarray}
\label{Result2}
N^r_{14}(m_{\rho})&=&2.8\cdot 10^{-2}, \nonumber \\
N^r_{15}(m_{\rho})&=&4.7\cdot 10^{-2}.
\end{eqnarray}
The results for the electric branching ratios obtained with these values 
for the counterterm couplings are listed in Table \ref{T+DAP}.
According to Tables \ref{T+L4} and \ref{T+L24}, one finds that $\mathcal{O}(p^4)$ corrections become 
more important for higher cuts in $q^2$ and that branching ratios for lower
cuts are dominated by the tree level. 
\begin{table}[h]
\caption{Total electric branching ratio of 
$K^+\rightarrow\pi^+\pi^0e^+e^-$ for cuts in $q^2$ and for
the entire phase space relying on the values of the LECs
calculated in (\ref{Result1}) and (\ref{Result2}). No errors are taken into 
account.}
\vspace*{0.2cm}
\hspace*{2.7cm}
\begin{tabular}{ll}
\hline\hline\noalign{\smallskip}
$q^2 > (\mbox{MeV}^2)$ & Electric $\mathcal{O}(p^2)+\mathcal{O}(p^4)$ 
BR $[10^{-9}]$\\ 
\noalign{\smallskip}\hline\noalign{\smallskip}
$2^2$ & $2923$\\ 
$10^2$ & $873$\\ 
$20^2$ & $431$\\ 
$30^2$ & $253$\\ 
$40^2$ & $164$\\ 
$60^2$ & $78$\\ 
$80^2$ & $39$\\ 
$100^2$ & $20$\\ 
$120^2$ & $9$\\ 
$180^2$ & $0.24$\\ 
$\mbox{entire p.s.}$ & $3683$\\ 
\noalign{\smallskip}\hline
\end{tabular}
\label{T+DAP}
\end{table}
E.g. for a cut of
$(10\;\mbox{MeV})^2$, the tree-level BR is modified by $\mathcal{O}(p^4)$
corrections and by the interference between the two electric contributions
by about $17\%$, whereas for a cut of $(80\;\mbox{MeV})^2$ the result is 
increased by roughly $160\%$. It is clear that the interference also gives
rise to an important part of the enhancement of the BR. Finally, one should 
note that the estimated couplings in (\ref{Result1}) and (\ref{Result2})
are in a range where one could expect them but it is also true that the 
uncertainties involved are too big to make a more precise statement about 
the couplings and the $K^+$ decay width.  
\section{Conclusions}
I considered the non-leptonic decays $K_L\rightarrow\pi^+\pi^-e^+e^-$ and
$K^+\rightarrow\pi^+\pi^0e^+e^-$ within the framework of Chiral
Perturbation Theory. First of all, the amplitudes of the decays have been 
given up to order $p^4$ in a very explicit way and a consistency check on
parts of the weak counter\-term Lagrangian of CHPT was performed:
all divergences are properly removed. 
\newline\hspace*{0.5cm}
The main reason to focus on $K_L\rightarrow\pi^+\pi^-e^+e^-$ in this
paper is provided by the possibility of extracting the combination 
$(N^r_{16}(\mu)-N_{17})$ of weak LECs from experimental results. 
The latest value of the preliminary branching ratio, 
$BR(K_L\rightarrow\pi^+\pi^-e^+e^-)=(3.63\pm 0.11\pm 0.14)
\cdot 10^{-7}$ \cite{BAR,KETT}, and
the values of the parameters of the magnetic form factor \cite{KTEV2},
both obtained by the KTeV collaboration, were used for the numerical
analysis of the decay.
\newline\hspace*{0.5cm}
The introduction of an energy dependent magnetic form factor yields an
important correction to the older calculations in \cite{SW,HS,ESW}, since
it increases the magnetic contribution to the branching ratio considerably.
The (preliminary) value for the branching ratio and the parameters
of the magnetic form factor were obtained from the analysis of the data set
of 1997 which contains more than 1800 events \cite{KTEV2}. Comparison with 
earlier experimental results shows that the errors of the measured 
quantities became quite smaller due to the better statistics but the 
uncertainties are still too big to make precise predictions.
\newline\hspace*{0.5cm}
Comparison with experiment yields two possible values of the LEC combination
$(N^r_{16}(m_{\rho})-N_{17})$, thus one has to consult theoretical
approaches about weak counterterm couplings to distinguish
between the possible solutions. All models that have been considered (Weak Deformation Model
WDM \cite{EKW}, Factorization Model FM \cite{EKW}, modified Factorization 
Model FMV \cite{DAP1,DAP}) prefer the same value of $(-1.4 \pm 3.6)
\cdot 10^{-2}$. Of course, the error, which is mainly due to 
the experimental uncertainties of the two parameters of the magnetic 
amplitude, is quite large, but nevertheless the obtained result is reasonable 
compared to $(N^r_{14}(m_{\rho})-N^r_{15}(m_{\rho}))=-1.9\cdot 10^{-2}$ and 
it rests upon a firm theoretical ground. 
\newline\hspace*{0.5cm}
On the other hand, the central value of $(N^r_{16}-N_{17})$ is almost in 
perfect agreement with the FM and WDM predictions. One also derives
central values for the two parameters of the FMV that are in good agreement
with the expectations. 
\newline\hspace*{0.5cm}
Since 1997, much more data have been collected by the KTeV group and 
therefore one can hope that a new analysis of the much bigger set of events can reduce the experimental
uncertainties. As already pointed out in Sect. 4.2, it should also be 
possible to extract the value of $(N^r_{16}-N_{17})$ to a better precision
by comparing the theoretical results with branching ratios for higher cuts in
$q^2$ (e.g. $\sim (40\;\mbox{MeV})^2$), since the contributions from loops and 
counterterms become much more important for higher cuts than for the entire 
phase space. 
\newline\hspace*{0.5cm} 
To be able to make a useful prediction for the $K^+$ decay, one has to rely
on additional theoretical assumptions. First, I followed the predictions of 
the FM and WDM and assumed that the extracted value $-1.4\cdot 10^{-2}$
is produced solely by $N^r_{16}$ and that $N_{17}=0$. It is therefore 
possible to express the branching ratio for
$K^+\rightarrow\pi^+\pi^0e^+e^-$ as a function of $(N^r_{14}+2N^r_{15})$. 
\newline\hspace*{0.5cm}
Contrary to the decay of the $K_L$, there exists an interference between the tree-level amplitude and the 
electric $\mathcal{O}(p^4)$ amplitude. In general, it is found that the 
branching ratio due to the electric part of the decay amplitude clearly 
dominates over the magnetic contributions. Moreover, it is the tree level 
that produces by far the most important contributions to the BR over the 
entire phase space as well as for a wide range of cuts in $q^2$. 
\newline\hspace*{0.5cm}
An extraction of the combination $(N_{14}^r+2N^r_{15})$ from the branching 
ratio over the entire phase space is almost impossible, but according to 
the discussion in Sect. 4.4, with a scan of the $q^2$ spectrum it is more 
likely to extract values for $(N^r_{14}+2N^r_{15})$, especially for cuts 
larger than $\sim (60\;\mbox{MeV})^2$. Obviously the experimental error of 
the magnetic 
part of the amplitude and the error of the combination $(N^r_{16}-N_{17})$ 
will not make it easier to extract a reasonable value, but hopefully new 
results from KTeV (and from CERN) also help to improve the predictive power 
of this analysis. 
\newline
\hspace*{0.5cm}
Whereas the analysis summarized so far was based on conservative
assumptions, I also speculated about extracting values for $N^r_{14}$,
$N^r_{15}$, $N^r_{16}$ and $N_{17}$. Referring to the FMV, the two parameters
of the model were estimated using the available data and the extracted
value of $(N^r_{16}-N_{17})$. The central values of these parameters were 
further used to estimate the central values of the four low-energy 
couplings $N^r_{14}$, $N^r_{15}$, $N^r_{16}$ and $N_{17}$ and to make a 
'prediction' of the $K^+$ branching ratio without any free parameter. 
Although the errors are big, the obtained values for the LECs seem to be 
reasonable.
\section*{Acknowledgements} 
Most of this work was done in Vienna and I would like to thank G. Ecker 
for his help through a long time. I thank G. Isidori, G. D'Ambrosio, 
J. Portol\'{e}s and R. Escri\-bano for helpful discussions on physics and 
computing. I am grateful for important comments on the experimental input by
T. Barker, B. Cox and S. Ledovskoy from the KTeV collaboration.
Finally, G. Pancheri should be mentioned since she made it possible for me
to come to the LNF in Frascati. 
\section*{Appendix}
\begin{appendix}
\section{Loop functions}
All loop integrals in this work can be reduced to a basis
of three scalar integrals:
\begin{eqnarray} 
\label{Basis}
iA(m^2)&=&\int \frac{d^dk}{(2\pi)^d}\frac{1}{D_1}, \nonumber \\
iB(q^2,m^2,M^2)&=&\int
\frac{d^dk}{(2\pi)^d}\frac{1}{D_2}, \nonumber \\
iC(q^2,p^2,m^2,M^2)&=&\int\frac{d^dk}{(2\pi)^d} 
\frac{1}{D_3},
\end{eqnarray}
where I introduced the abbreviations $D_1=[k^2-m^2]$, 
$D_2=[k^2-m^2][(k-q)^2-M^2]$ and
$D_3=[k^2-m^2][(k-q)^2-M^2][(k-p)^2-M^2]$. 
Indexed loop functions are defined through the following relations:
\begin{eqnarray}
\int\frac{d^dk}{(2\pi)^d}\frac{k_{\mu}}{D_2}&=&
iq_{\mu}B_1(q^2,m^2,M^2), \\
\int\frac{d^dk}{(2\pi)^d}\frac{k_{\mu}}{D_3}&=&
iq_{\mu}C_1(q^2,p^2,qp,m^2,M^2)+ip_{\mu}C_2(q^2,p^2,qp,m^2,M^2),\nonumber \\
\int\frac{d^dk}{(2\pi)^d}\frac{k_{\mu}k_{\nu}}{D_2}&=&
ig_{\mu\nu}B_{00}(q^2,m^2,M^2)+iq_{\mu}q_{\nu}B_{11}(q^2,m^2,M^2),\nonumber \\
\int\frac{d^dk}{(2\pi)^2}\frac{k_{\mu}k_{\nu}}{D_3}&=&
ig_{\mu\nu}C_{00}(q^2,p^2,qp,m^2,M^2)+iq_{\mu}q_{\nu}
C_{11}(q^2,p^2,qp,m^2,M^2)\nonumber \\
& &\hspace{-2.6cm}+i(q_{\mu}p_{\nu}+q_{\nu}p_{\mu})C_{12}(q^2,p^2,qp,m^2,M^2)+
ip_{\mu}p_{\nu}C_{22}(q^2,p^2,qp,m^2,M^2).\nonumber
\end{eqnarray}
They can be given explicitly in terms of (\ref{Basis}). Divergences arise 
through the scalar loop functions $A$ and $B$ in (\ref{Basis}). The
divergent parts of these functions are isolated in expressions similar 
to (\ref{LEC}).
\begin{eqnarray}
&&A(m^2)|_{div}=-2m^2\Lambda(\mu), \quad B(q^2,m^2,M^2)|_{div}=
-2\Lambda(\mu), \nonumber \\
&&\Lambda(\mu)=\frac{\mu^{d-4}}{16\pi^2}\Big
[\frac{1}{d-4}-\frac{1}{2}(\ln (4\pi) +1-\gamma_E)\Big ].
\end{eqnarray}
Apart from $C_{00}$, all $C$-like functions are finite.
\section{\boldmath $K_L \rightarrow \pi^+\pi^-\gamma^{\star}$ form factor 
\boldmath $\mathcal{F}^{Ll}_{16}$}
The electric form factor $\mathcal{F}^L_1$ gets contributions from topologies
2 and 3 collected in $\mathcal{F}^{Ll}_{16}$; the first part of 
$\mathcal{F}^{Ll}_{16}$ is due to the loop particles $(K^0_1,\pi^{\pm})$, the 
second part arises from $(\eta_8,K^{\pm})$, and the very last line comes from 
$(\pi^0,K^{\pm})$. 
\begin{eqnarray}
\label{FF1B}
\mathcal{F}_{16}^{Ll} & = & \frac{-ieG_8}{F}\cdot \bigg
\{-2A(m_{\pi}^2)-\frac{1}{2}(q^2+2qp_1)B(q^2,m_{\pi}^2,m_{\pi}^2)+\frac{1}{2(q^2+2qp_1)}\cdot
\nonumber \\
& &\Big[2m_{\pi}^2(q^2+2qp_1-m_K^2)+m_K^2(m_K^2-q^2-2qp_1)\Big ]
\Big (B(m_{\pi}^2,m_K^2,m_{\pi}^2)\nonumber \\
& &-B((p_1+q)^2,m_K^2,m_{\pi}^2)\Big )+\frac{1}{q^2+2qp_1}\Big [m_K^2
(m_{\pi}^2+p_1p_2+qp_2-q^2\nonumber \\
& &-2qp_1)-2m_{\pi}^2(p_1p_2+qp_2)\Big]
B_1(m_{\pi}^2,m_K^2,m_{\pi}^2)+\frac{1}{2(q^2+2qp_1)}\Big [m_K^2(-2m_{\pi}^2\nonumber \\
& &-q^2-2qp_1-2p_1p_2-2qp_2)+m_{\pi}^2(2q^2+4qp_1+4p_1p_2+4qp_2)+2q^4\nonumber
\\&&+q^2(8qp_1+2p_1p_2+2qp_2)+8qp_1^2+4p_1p_2qp_1+4qp_2qp_1\Big ]\nonumber \\
&&B_1((p_1+q)^2,m_K^2,m_{\pi}^2)+4B_{00}(q^2,m_{\pi}^2,m_{\pi}^2)+
\frac{2(p_1p_2+qp_2)}{q^2+2qp_1}\nonumber \\
&&\Big[B_{00}(m_{\pi}^2,m_K^2,m_{\pi}^2)
-B_{00}((p_1+q)^2,m_K^2,m_{\pi}^2)+m_{\pi}^2B_{11}(m_{\pi}^2,m_K^2,m_{\pi}^2)
\nonumber \\
&&-(m_{\pi}^2+q^2+2qp_1)B_{11}((p_1+q)^2,m_K^2,m_{\pi}^2)\Big ]-
\frac{1}{2}\Big[m_K^2(m_K^2-q^2-2qp_1\nonumber \\
&& -2m_{\pi}^2)+2m_{\pi}^2(q^2+2qp_1)\Big]C(m_{\pi}^2,(p_1+q)^2,m_K^2,m_{\pi}
^2)+\frac{1}{2}\Big[m_K^2(-4p_1p_2\nonumber \\
&&-4m_{\pi}^2+m_K^2-2q^2-4qp_1)+m_{\pi}^2(4p_1p_2+4q^2+8qp_1)\Big]\nonumber\\
& &C_1(m_{\pi}^2,(p_1+q)^2,m_K^2,m_{\pi}^2)
+\frac{1}{2}\Big [m_K^2(-4m_{\pi}^2-6qp_1-4p_1p_2-4qp_2+
m_K^2\nonumber \\
&&-2q^2)+4m_{\pi}^2(q^2+2qp_1+p_1p_2+qp_2)+2qp_1(q^2+2qp_1)\Big ]\nonumber \\
&&C_2(m_{\pi}^2,(p_1+q)^2,m_K^2,m_{\pi}^2)+m_K^2C_{00}(m_{\pi}^2,
(p_1+q)^2,m_K^2,m_{\pi}^2)-2(m_{\pi}^2\nonumber \\
&&-m_K^2)C_{00}(m_{\pi}^2,(p_2+q)^2,m_K^2,m_{\pi}^2)+(2m_K^2p_1p_2+
m_K^2m_{\pi}^2-2m_{\pi}^2p_1p_2)\nonumber \\
&&C_{11}(m_{\pi}^2,(p_1+q)^2,m_K^2,m_{\pi}^2)+\Big[m_K^2(qp_1+2qp_2+2m_{\pi}^2
+4p_1p_2)-2m_{\pi}^2(qp_2\nonumber \\
&&+2p_1p_2)\Big ]C_{12}(m_{\pi}^2,(p_1+q)^2,m_K^2,m_{\pi}^2)
+\Big[m_K^2(2p_1p_2+2qp_2+m_{\pi}^2+qp_1)\nonumber \\
&&-2m_{\pi}^2
(p_1p_2+qp_2)\Big ]C_{22}(m_{\pi}^2,(p_1+q)^2,m_K^2,m_{\pi}^2)\nonumber \\
& &\hspace{-0.5cm}\nonumber \\
&&-\frac{11}{6}A(m_K^2)-\frac{1}{9}\Big[m_{\pi}^2+2m_K^2+6
(2qp_1+qp_2+p_1p_2+q^2)\Big ]B(q^2,m_K^2,m_K^2)\nonumber\\
& &+\frac{1}{18(q^2+2qp_1)}\Big [m_{\eta}^2\Big(-6(p_1p_2+qp_2+
q^2+2qp_1)+2m_K^2-11m_{\pi}^2\Big)+3m_{\pi}^2\nonumber\\
&&(m_{\pi}^2+6q^2+12qp_1+2m_K^2+6p_1p_2+
6qp_2)\Big ]\Big (B(m_{\pi}^2,m_{\eta}^2,m_K^2)\nonumber \\
& &-B((p_1+q)^2,m_{\eta}^2,m_K^2)\Big)+\frac{1}{9(q^2+2qp_1)}
\Big[9m_{\eta}^2(m_{\pi}^2+p_1p_2+qp_2-q^2-2qp_1)\nonumber \\
& &+m_{\pi}^2(-11m_{\pi}^2-9q^2
-18qp_1-4m_K^2-39p_1p_2-39qp_2)+6m_K^2(q^2+2qp_1)\Big]\nonumber \\
& &B_1(m_{\pi}^2,m_{\eta}^2,m_K^2)-\frac{1}{9(q^2+2qp_1)}
\Big [3m_{\eta}^2(2q^2+4qp_1+3p_1p_2+3qp_2+3m_{\pi}^2)\nonumber \\
& &+m_{\pi}^2(-11m_{\pi}^2
-4m_K^2-29q^2-58qp_1-39p_1p_2-39qp_2)-4m_K^2(q^2+2qp_1)\nonumber \\
& &-72qp_1^2-72q^2qp_1
-18q^4-30q^2p_1p_2-30q^2qp_2-60qp_1p_1p_2-60qp_1qp_2\Big ]\nonumber \\
& &B_1((p_1+q)^2,m_{\eta}^2,m_K^2)+5B_{00}(q^2,m_K^2,m_K^2)+
\frac{4}{3(q^2+2qp_1)}\Big[(m_{\pi}^2+3p_1p_2\nonumber \\
&&+3qp_2)B_{00}(m_{\pi}^2,m_{\eta}^2,
m_K^2)-(m_{\pi}^2+q^2+2qp_1+3p_1p_2+3qp_2)\nonumber\\
& &B_{00}((p_1+q)^2,m_{\eta}^2,m_K^2)
\Big ]+\frac{4}{3(q^2+2qp_1)}\Big[m_{\pi}^2(m_{\pi}^2+3p_1p_2+3qp_2)\nonumber \\
& &B_{11}(m_{\pi}^2,m_{\eta}^2,m_K^2)-\Big (m_{\pi}^2(m^2_{\pi}+2q^2+4qp_1+3p_1p_2
+3qp_2)+q^2(q^2+4qp_1\nonumber \\
&&+3p_1p_2+3qp_2)+4qp_1^2+6qp_1(qp_2+p_1p_2)\Big )B_{11}((p_1+q)^2,
m_{\eta}^2,m_K^2)\Big ]\nonumber \\
&&-\frac{1}{18}
\Big [m_{\eta}^2\Big(-6(p_1p_2+q^2+2qp_1+qp_2)-11m_{\pi}^2+2m_K^2\Big )
+3m_{\pi}^2(6p_1p_2\nonumber \\
&&+6qp_2+6q^2+12qp_1+m_{\pi}^2+2m_K^2)\Big]C(m_{\pi}^2,(p_1+q)^2,
m_{\eta}^2,m_K^2)+\frac{1}{18}\Big [3m_{\eta}^2\nonumber \\
&&(-14m_{\pi}^2-24p_1p_2-18qp_1-6qp_2-6q^2-2m_K^2+m_{\eta}^2
)+m_{\pi}^2(24m_K^2\nonumber \\
&&+30qp_2+66qp_1+48p_1p_2+30q^2+11m_{\pi}^2)+m_K^2(36qp_1+12qp_2+12q^2\nonumber
\\
&&+48p_1p_2+4m_K^2)\Big]C_1(m_{\pi}^2,(p_1+q)^2,m_{\eta}^2,m_K^2)+\frac{1}{18}
\Big [3m_{\eta}^2(m_{\eta}^2-2m_K^2\nonumber \\
&&-14m_{\pi}^2-12q^2-28qp_1-24qp_2-24p_1p_2)
+m_{\pi}^2(11m_{\pi}^2+24m_K^2+36q^2\nonumber \\
&&+76qp_1+48qp_2+48p_1p_2)+4m_K^2(6q^2+14qp_1+12qp_2+12p_1p_2+m_K^2)\nonumber
\\
&&+24qp_1(q^2+qp_2+p_1p_2+2qp_1)\Big ]C_2(m_{\pi}^2,(p_1+q)^2,m_{\eta}^2,m_K^2)+
2(m_K^2-m_{\pi}^2)\nonumber \\
& &C_{00}(m_{\pi}^2,(p_2+q)^2,m_{\eta}^2,m_K^2)+\frac{2}{3}
(m_{\eta}^2+m_K^2-m_{\pi}^2)C_{00}(m_{\pi}^2,(p_1+q)^2,m_{\eta}^2,m_K^2)
\nonumber \\
&&+\frac{1}{3}\Big [m_{\eta}^2(5m_{\pi}^2+3qp_1+9p_1p_2)-m_{\pi}^2(m_{\pi}^2+2m_K^2+qp_1+3p_1p_2)-2m_K^2\nonumber \\
&&(qp_1+3p_1p_2)\Big ]C_{11}(m_{\pi}^2,(p_1+q)^2,m_{\eta}^2,m_K^2)+\frac{1}{3}
\Big [m_{\eta}^2(10m_{\pi}^2+3q^2+11qp_1\nonumber \\
&&+9qp_2+18p_1p_2)-m_{\pi}^2(2m_{\pi}^2+4m_K^2+q^2+3qp_1+3qp_2
+6p_1p_2)-2m_K^2(q^2\nonumber \\
&&+3qp_1+3qp_2+6p_1p_2)\Big ]C_{12}(m_{\pi}^2,(p_1+q)^2,m_{\eta}^2,m_K^2)
+\frac{1}{3}\Big [m_{\eta}^2(5m_{\pi}^2+3q^2\nonumber \\
&&+8qp_1+9qp_2+9p_1p_2)-m_{\pi}^2(m_{\pi}^2+2m_K^2+q^2+2qp_1+3qp_2+3p_1p_2)
\nonumber \\
&&-2m_K^2(q^2+2qp_1+3qp_2+3p_1p_2)\Big]
C_{22}(m_{\pi}^2,(p_1+q)^2,m_{\eta}^2,m_K^2)\nonumber \\
& &+\frac{(d-2)}{2(d-1)}A(m_K^2)+\frac{1}{4(d-1)}
B(q^2,m_K^2,m_K^2)(q^2-4m_K^2)\bigg \}.
\end{eqnarray}
\section{\boldmath $K^+\rightarrow \pi^+\pi^0 \gamma^{\star}$ form factors
\boldmath $\mathcal{F}^{+l}_{15,16}$, \boldmath $\mathcal{F}^{+l}_{25,26}$}
The electric form factors $\mathcal{F}^+_1$ and $\mathcal{F}^+_2$ get 
contributions from topologies 2 and 3. Contributions with an $\eta_8$ in 
the loop are collected in expressions $\mathcal{F}^{+l}_{15,25}$, 
contributions with a pair of any kaon and pion are collected in 
$\mathcal{F}^{+l}_{16,26}$, respectively. The contributions with an 
$\eta_8$ read as:
\begin{eqnarray}
\label{FF+l15}
\mathcal{F}^{+l}_{15}&=&\frac{-ieG_8}{F}\cdot \bigg \{-\frac{1}{9(q^2+2qp_1)(q^2+2qp_1+
2qp_2)}\Big [m_{\pi}^2(6q^2+12qp_1+22qp_2)\nonumber \\
& &+m_K^2(-3q^2-6qp_1-qp_2)+6q^2(q^2+
4qp_1+4qp_2+p_1p_2)+12qp_1(2qp_1\nonumber \\
& &+4qp_2+p_1p_2)+24qp_2(qp_2+p_1p_2)\Big ]
A(m_K^2)+\frac{1}{18(q^2+2qp_1)(q^2+2qp_1+2qp_2)}\nonumber \\
&&\Big[3m_{\eta}^2(-2m_{\pi}^2+
m_K^2-q^2-2qp_1-2p_1p_2)+m_{\pi}^2(26m_{\pi}^2-15m_K^2+13q^2\nonumber \\
&&+26qp_1+12qp_2+
38p_1p_2)+m_K^2(m_K^2-q^2-2qp_1-6qp_2-8p_1p_2)+6p_1p_2\nonumber \\
&&(q^2+2qp_1+2qp_2+2p_1p_2)\Big ]A(m_{\eta}^2)+\frac{1}{9}
(11m_{\pi}^2-8m_K^2+6qp_2+6p_1p_2)\nonumber \\
&&B(q^2,m_K^2,m_K^2)+\frac{1}
{54(q^2+2qp_1)(q^2+2qp_1+2qp_2)}\Big[m_{\pi}^2m_K^2(-12m_{\pi}^2+24m_K^2
\nonumber \\
&&-58q^2-116qp_1-60p_1p_2)+m_{\pi}^2p_1p_2(-12q^2-24qp_1-240m_{\pi}^2-
144p_1p_2\nonumber \\
&&-144qp_2)+m_K^2p_1p_2(30m_K^2-60q^2-120qp_1-72qp_2-72p_1p_2)+m_{\pi}^4
\nonumber \\
&&(-28q^2-56qp_1-144qp_2)+m_K^4(5q^2+10qp_1+36qp_2)+m_{\pi}^2(60q^4+
240qp_1^2\nonumber \\
&&+240q^2qp_1+120q^2qp_2+240qp_1qp_2)+m_K^2(-24q^4-96qp_1^2-96q^2qp_1
\nonumber \\
&&-48q^2qp_2-96qp_1qp_2)+3m_K^6-96m_{\pi}^6\Big ]B(m_{\pi}^2,m_{\eta}^2,m_K^2)+
\frac{1}{18(q^2+2qp_1)}\nonumber \\
&&\Big [m_{\eta}^2(-3m_{\eta}^2+8m_K^2-2m_{\pi}^2-6qp_2-6p_1p_2)+m_{\pi}^2
(33m_{\pi}^2-24m_K^2+18qp_2\nonumber \\
&&+18p_1p_2)\Big ]
B((p_1+q)^2,m_{\eta}^2,m_K^2)-\frac{1}{18(q^2+2qp_1+2qp_2)}\Big
[9m_{\eta}^2(-2m_{\pi}^2-q^2\nonumber \\
&&-2qp_1-2qp_2-2p_1p_2)+m_{\pi}^2(22m_{\pi}^2+8m_K^2+27q^2+54qp_1+54qp_2\nonumber
\\
&&+78p_1p_2)+6m_K^2(q^2+2qp_1+2qp_2)\Big ]B_1(m_{\pi}^2,m_{\eta}^2,m_K^2)
\nonumber \\
&&+\frac{1}{(q^2+2qp_1)(q^2+2qp_1+2qp_2)}\Big [m_{\pi}^2m_K^2(-2m_{\pi}^2-m_K^2+q^2
+2qp_1-4qp_2\nonumber \\
&&-2p_1p_2)+m_{\pi}^2p_1p_2(20m_{\pi}^2+6q^2+12qp_1+
12qp_2+12p_1p_2)+m_K^2p_1p_2\nonumber \\
&&(-m_K^2+q^2+2qp_1+2qp_2+2p_1p_2)+4m^4(q^2+2qp_1+3qp_2+
2m_{\pi}^2)-m_K^4qp_2\Big]\nonumber \\
&&B_1(m_{\pi}^2,m_K^2,m_{\eta}^2)+\frac{1}{9(q^2+2qp_1)}
\Big [m_{\pi}^2(-10m_{\pi}^2+4m_K^2-5q^2-10qp_1+18qp_2\nonumber \\
&&+18p_1p_2)+4m_K^2(2q^2+4qp_1-3qp_2-3p_1p_2)+6q^2(q^2+4qp_1+qp_2\nonumber
\\
&&+p_1p_2)+12qp_1(2qp_1+qp_2+p_1p_2)\Big ]
B_1((p_1+q)^2,m_{\eta}^2,m_K^2)\nonumber \\
& &+2B_{00}(q^2,m_K^2,m_K^2)+\frac{4(m_{\pi}^2+
3p_1p_2)}{3(q^2+2qp_1+2qp_2)}B_{00}(m_{\pi}^2,m_{\eta}^2,m_K^2)\nonumber
\\&&-\frac{2}{(q^2+2qp_1)(q^2+2qp_1+2qp_2)}\Big [2m_{\pi}^2(2m_{\pi}^2-m_K^2+q^2+2qp_1
+3qp_2\nonumber \\
&&+5p_1p_2)-3m_K^2(qp_2+p_1p_2)+3p_1p_2(q^2+2qp_1+2qp_2+2p_1p_2)\Big ]
\nonumber \\
&&B_{00}(m_{\pi}^2,m_K^2,m_{\eta}^2)-\frac{2}{3(q^2+2qp_1)}\Big [2m_{\pi}^2+6qp_2+
6p_1p_2+2q^2+4qp_1\Big ]\nonumber \\
&&B_{00}((p_1+q)^2,m_{\eta}^2,m_K^2)+\frac{4m_{\pi}^2(m_{\pi}^2+3p_1p_2)}
{3(q^2+2qp_1+2qp_2)}B_{11}(m_{\pi}^2,m_{\eta}^2,m_K^2)\nonumber \\
&&-\frac{2m_{\pi}^2}{(q^2+2qp_1)(q^2+2qp_1+2qp_2)}\Big
[2m_{\pi}^2(2m_{\pi}^2
-m_K^2+q^2+2qp_1+3qp_2\nonumber \\
&&+5p_1p_2)-3m_K^2(qp_2+p_1p_2)+3p_1p_2(q^2+2qp_1+2qp_2+2p_1p_2)\Big ]
\nonumber \\
&&B_{11}(m_{\pi}^2,m_K^2,m_{\eta}^2)-\frac{4}{3(q^2+2qp_1)}\Big [m_{\pi}^2
(m_{\pi}^2+2q^2+4qp_1+3qp_2+3p_1p_2)\nonumber \\
&&+q^2(q^2+4qp_1+3qp_2+3p_1p_2)+qp_1(4qp_1+6qp_2+6p_1p_2)\Big ]\nonumber \\
&&B_{11}((p_1+q)^2,m_{\eta}^2,m_K^2)+\frac{1}{18}\Big [\Big (
m_{\eta}^2(-3m_{\eta}^2-2m_{\pi}^2+8m_K^2-6qp_2-6p_1p_2)\nonumber \\
&&+m_{\pi}^2(33m_{\pi}^2-24m_K^2+
18qp_2+18p_1p_2)\Big )C(m_{\pi}^2,(p_1+q)^2,m_{\eta}^2,m_K^2)\nonumber \\
& &-\Big(m_{\eta}^2(-3m_{\eta}^2+6m_{\pi}^2+24m_K^2+18qp_1-18qp_2+
36p_1p_2)+m_{\pi}^2(49m_{\pi}^2\nonumber \\
&&-30m_K^2-6qp_1+30qp_2+12p_1p_2)+m_K^2(-16m_K^2
-12qp_1+12qp_2\nonumber \\
&&-24p_1p_2)\Big )C_1(m_{\pi}^2,(p_1+q)^2,m_{\eta}^2,m_K^2)
-\Big(m_{\eta}^2(-3m_{\eta}^2+6m_{\pi}^2+24m_K^2\nonumber \\
&&+18q^2+48qp_1+36qp_2+36p_1p_2)+m_{\pi}^2(49m_{\pi}^2-30m_K^2-6q^2+32qp_1\nonumber
\\
&&+12qp_2+12p_1p_2)+m_K^2(-16m_K^2-12q^2-56qp_1-24qp_2-24p_1p_2)
\nonumber \\
&&+24qp_1qp_2+24qp_1p_1p_2\Big)C_2(m_{\pi}^2,(p_1+q)^2,
m_{\eta}^2,m_K^2)\Big ]-\frac{1}{3}\Big [(-5m_{\eta}^2+m_{\pi}^2\nonumber
\\
&&+2m_K^2)C_{00}(m_{\pi}^2, 
(p_1+q)^2,m_{\eta}^2,m_K^2)+\Big (m_{\eta}^2(-5m_{\pi}^2-3qp_1-9p_1p_2)+
m_{\pi}^2\nonumber \\
&&(m_{\pi}^2+2m_K^2+qp_1+3p_1p_2)+m_K^2(2qp_1+6p_1p_2)\Big)\nonumber \\
& &C_{11}(m_{\pi}^2,(p_1+q)^2,m_{\eta}^2,m_K^2)+
\Big(m_{\eta}^2(-5m_{\pi}^2-3q^2-8qp_1
-9qp_2-9p_1p_2)\nonumber \\
& &+m_{\pi}^2(m_{\pi}^2+2m_K^2+q^2+2qp_1+3qp_2+3p_1p_2)+m_K^2(
2q^2+4qp_1+6qp_2\nonumber \\
&&+6p_1p_2)\Big )C_{22}(m_{\pi}^2,(p_1+q)^2,m_{\eta}^2,m_K^2)+
\Big (m_{\eta}^2(-10m_{\pi}^2-3q^2-11qp_1\nonumber \\
&&-9qp_2-18p_1p_2)+m_{\pi}^2(2m_{\pi}^2
+4m_K^2+q^2+3qp_1+3qp_2+6p_1p_2)\nonumber \\
& &+m_K^2(2q^2+6qp_1+6qp_2+12p_1p_2)\Big )C_{12}(m_{\pi}^2,
(p_1+q)^2,m_{\eta}^2,m_K^2)\Big ]\bigg \},
\end{eqnarray}
\begin{eqnarray}
\label{FF+l25}
\mathcal{F}^{+l}_{25}&=&\frac{-ieG_8}{F}\cdot \bigg
\{\frac{1}{18(q^2+2qp_1+2qp_2)}(-3q^2-6qp_1-6qp_2+12p_1p_2+5m_K^2\nonumber \\
& &+10m_{\pi}^2)A(m_K^2)
+\frac{1}{18(q^2+2qp_1+2qp_2)}(54p_1p_2+62m_{\pi}^2-29m_K^2)A(m_{\eta}^2)
\nonumber\\& &-\frac{1}{54(q^2+2qp_1+2qp_2)}\Big[m_{\pi}^2(-30q^2-60qp_1-
60qp_2-60p_1p_2-44m_{\pi}^2\nonumber \\
& &+58m_K^2)+m_K^2(12q^2+24qp_1+24qp_2+24p_1p_2+13m_K^2)\Big]
B(m_{\pi}^2,m_{\eta}^2,m_K^2)\nonumber \\
& &-\frac{1}{9(q^2+2qp_1+2qp_2)}\Big[m_{\pi}^2(6q^2+12qp_1+12qp_2+
39p_1p_2+11m_{\pi}^2+4m_K^2\nonumber \\
& &-9m_{\eta}^2)-9p_1p_2m_{\eta}^2\Big]B_1(m_{\pi}^2,m_{\eta}^2
,m_K^2)+\frac{1}{2(q^2+2qp_1+2qp_2)}\Big[m_{\pi}^2(2m_{\pi}^2\nonumber \\
& &+6p_1p_2+9m_K^2-6m_{\eta}^2)+m_K^2(-2m_K^2+6p_1p_2+3m_{\eta}^2)-
6p_1p_2m_{\eta}^2\Big ]\nonumber \\
& &B_1(m_{\pi}^2,m_K^2,m_{\eta}^2)+3B_{00}(q^2,m_K^2,m_K^2)-2B_{00}((p_1+q)^2,
m_{\eta}^2,m_K^2)\nonumber \\
& &+\frac{4(m_{\pi}^2+3p_1p_2)}{3(q^2+2qp_1+2qp_2)}B_{00}
(m_{\pi}^2,m_{\eta}^2,m_K^2)+2B_{00}((p_1+q)^2,m_{\eta}^2,m_K^2)\nonumber\\
& &-\frac{2(2m_{\pi}^2+
3p_1p_2)}{q^2+2qp_1+2qp_2}B_{00}(m_{\pi}^2,m_K^2,m_{\eta}^2)+\frac{4m_{\pi}^2
(m_{\pi}^2+3p_1p_2)}{3(q^2+2qp_1+2qp_2)}\nonumber \\
&& B_{11}(m_{\pi}^2,m_{\eta}^2,m_K^2)-\frac{2m_{\pi}^2(2m_{\pi}^2+3p_1p_2)}
{q^2+2qp_1+2qp_2}B_{11}(m_{\pi}^2,m_K^2,m_{\eta}^2)+2(m_K^2-m_{\pi}^2)
\nonumber \\ 
&&C_{00}(m_{\pi}^2,(p_1+q)^2,m_{\eta}^2,m_K^2)\bigg \}.
\end{eqnarray}
\hspace*{0.5cm}
The remaining contributions from $(K,\pi)$ pairs are collected in the 
form factors $\mathcal{F}^{+l}_{16}$ and $\mathcal{F}^{+l}_{26}$. The first 
parts refer to $(K^0,\pi^0)$ contributions, then $(\pi^0,K^-)$,
$(K^0,\pi^-)$ and $(K^+,\pi^+)$ contributions are presented.
\begin{eqnarray} 
\label{FF+l16}
\mathcal{F}^{+l}_{16}&=&\frac{-ieG_8}{F}\bigg\{\frac{1}{(q^2+2qp_1)(q^2+2qp_1+2qp_2)}\Big [
-\frac{1}{2}\Big (2m_{\pi}^2(2m_{\pi}^2+q^2+2qp_1\nonumber \\
& &+2qp_2+4p_1p_2)+m_K^2(-m_K^2+q^2+
2qp_1-2qp_2)+2p_1p_2(q^2\nonumber \\
&&+2qp_1+2qp_2+2p_1p_2)\Big)A(m_{\pi}^2)+\frac{m_K^2}{2}\Big (
2m_{\pi}^2(2m_{\pi}^2-2m_K^2\nonumber \\
&&+q^2+2qp_1+2p_1p_2)+m_K^2(m_K^2-q^2-2qp_1-2p_1p_2)\Big )\nonumber \\
&&B(m_{\pi}^2,m_K^2,m_{\pi}^2)-\Big (m_{\pi}^2m_K^2(-m_K^2+2m_{\pi}^2+q^2
+2qp_1+4qp_2+6p_1p_2)\nonumber \\
&&-2m_{\pi}^2p_1p_2(
2m_{\pi}^2+q^2+2qp_1+2qp_2+2p_1p_2)+m_K^2p_1p_2(-m_K^2+q^2\nonumber \\
&&+2qp_1+2qp_2+2p_1p_2)-
4m_{\pi}^4qp_2-m_K^4qp_2\Big )B_1(m_{\pi}^2,m_K^2,m_{\pi}^2)\nonumber \\
&&-2\Big (p_1p_2(
2m_{\pi}^2-m_K^2+q^2+2qp_1+2qp_2+2p_1p_2)+2m_{\pi}^2qp_2-m_K^2qp_2\Big )
\nonumber \\
&&\Big(B_{00}(m_{\pi}^2,m_K^2,m_{\pi}^2)+m_{\pi}^2B_{11}(m_{\pi}^2,m_K^2,m_{\pi}^2)\Big )
\Big ]\nonumber \\
& &+\frac{1}{3(q^2+2qp_1)(q^2+2qp_1+2qp_2)}\Big [2m_{\pi}^2(q^2+2qp_1+
4qp_2)+m_K^2\nonumber \\
&&(-q^2-2qp_1+qp_2)+2q^2(q^2+4qp_1+4qp_2+p_1p_2)+4p_1p_2(qp_1\nonumber \\
&&+2qp_2)+8(qp_1^2+qp_2^2+2qp_1qp_2)\Big
]A(m_K^2)-\frac{1}{3(d-1)}\nonumber \\
&&A(m_K^2)+\frac{1}{6(d-1)}
\Big (q^2-4m_K^2\Big )B(q^2,m_K^2,m_K^2)\nonumber \\
& &-\frac{1}{(q^2+2qp_1)(q^2+2qp_1+2qp_2)}\Big [2m_{\pi}^2(q^2+2qp_1+
3qp_2)-m_K^2(q^2\nonumber \\
&&+2qp_1)+2q^2(q^2+4qp_1+4qp_2+p_1p_2)+4p_1p_2(qp_1+2qp_2)+8(qp_1^2\nonumber
\\
&&+qp_2^2+2qp_1qp_2)\Big ]A(m_{\pi}^2)+\frac{1}{q^2+2qp_1+2qp_2}\Big [-\Big
(q^2(m_K^2-m_{\pi}^2)\nonumber \\
&&-m_K^4+m_{\pi}^2m_K^2-2m_{\pi}^2(qp_1+qp_2)+2m_K^2(qp_1+qp_2)
\Big)B(m_{\pi}^2,m_K^2,m_{\pi}^2)\nonumber \\
& &+2p_1p_2(m_K^2-m_{\pi}^2)B_1(m_{\pi}^2,m_K^2,
m_{\pi}^2)\Big ]-\frac{1}{q^2+2qp_1}\Big [\Big
(-2m_{\pi}^4+2m_{\pi}^2m_K^2\nonumber \\
& &-2m_{\pi}^2
(qp_2+p_1p_2)+2m_K^2(qp_2+p_1p_2)\Big
)B((p_1+q)^2,m_K^2,m_{\pi}^2)\nonumber \\
&&+2(m_K^2-m_{\pi}^2)(qp_2+p_1p_2)B_1((p_1+q)^2,m_K^2,m_{\pi}^2)\Big ]
\nonumber \\
&&+2B_{00}(q^2,m_{\pi}^2,m_{\pi}^2)+\Big[2m_{\pi}^2(m_{\pi}^2-m_K^2+qp_2
+p_1p_2)-2m_K^2(qp_2\nonumber \\
&&+p_1p_2)\Big ]C(m_{\pi}^2,(p_1+q)^2,m_K^2,m_{\pi}^2)-\Big [2m_{\pi}^2
(m_{\pi}^2-m_K^2+qp_2)\nonumber \\
&&-2m_K^2qp_2\Big ]C_1(m_{\pi}^2,(p_1+q)^2,m_K^2,m_{\pi}^2)-
2m_{\pi}^2(m_{\pi}^2-m_K^2)\nonumber \\
& &C_2(m_{\pi}^2,(p_1+q)^2,m_K^2,m_{\pi}^2)-
2p_1p_2(m_{\pi}^2-m_K^2)\nonumber \\
&&C_{11}(m_{\pi}^2,(p_1+q)^2,m_K^2,m_{\pi}^2)-2(m_{\pi}^2-m_K^2)
(qp_2+p_1p_2)\nonumber \\
&&C_{22}(m_{\pi}^2,(p_1+q)^2,m_K^2,m_{\pi}^2)-2(m_{\pi}^2-m_K^2)(qp_2+2p_1p_2)
\nonumber \\
&&C_{12}(m_{\pi}^2,(p_1+q)^2,m_K^2,m_{\pi}^2)\nonumber\\
& &+\frac{1}{3(q^2+2qp_1)(q^2+2qp_1+2qp_2)}\Big [-\Big (2m_{\pi}^2
(4m_{\pi}^2-4m_K^2-q^2-2qp_1\nonumber \\
&&+8qp_2+12p_1p_2)+m_K^2(2m_K^2+q^2+2qp_1-8qp_2-12p_1p_2)+
2p_1p_2\nonumber \\
&&(q^2+2qp_1+8qp_2+8p_1p_2)\Big )A(m_{\pi}^2)-m_K^2\Big (2m_{\pi}^2(2m_{\pi}^2-
3m_K^2-2q^2\nonumber \\
&&-4qp_1+2qp_2+4p_1p_2)+m_K^2(2m_K^2+q^2+2qp_1-2qp_2-6p_1p_2)\nonumber \\
&&+4p_1p_2(-q^2-2qp_1+qp_2+p_1p_2)\Big )B(m_{\pi}^2,m_K^2,m_{\pi}^2)+2\Big
(4m_{\pi}^6\nonumber \\
&&+4m_{\pi}^4(-2m_K^2-q^2-2qp_1+qp_2+2p_1p_2)+3m_K^4(m_{\pi}^2+qp_2+p_1p_2)
\nonumber \\
&&+2m_{\pi}^2p_1p_2
(-7m_K^2-2q^2-4qp_1+2qp_2+2p_1p_2)-3m_K^2p_1p_2(q^2\nonumber \\
&&+2qp_1+2qp_2+2p_1p_2)-8m_{\pi}^2
m_K^2qp_2\Big )B_1(m_{\pi}^2,m_K^2,m_{\pi}^2)+4\Big (m_{\pi}^2\nonumber \\
&&(2m_{\pi}^2-m_K^2+q^2+
2qp_1+6qp_2+8p_1p_2)-3m_K^2(qp_2+p_1p_2)\nonumber \\
&&+3p_1p_2(q^2+2qp_1+2qp_2+2p_1p_2)\Big )
\Big (B_{00}(m_{\pi}^2,m_K^2,m_{\pi}^2)+m_{\pi}^2\nonumber \\
&&B_{11}(m_{\pi}^2,m_K^2,m_{\pi}^2)\Big )\Big ]+2B_{00}(q^2,m_K^2,m_K^2)\bigg \}.
\end{eqnarray}
$\mathcal{F}^{+l}_{26}$ is organized in the same way and reads as follows:
\begin{eqnarray}
\label{FF+l26}
\mathcal{F}^{+l}_{26}&=&\frac{-ieG_8}{F}\bigg \{\frac{1}{q^2+2qp_1+2qp_2}\Big [
\frac{1}{2}(2m_{\pi}^2-3m_K^2+2p_1p_2)A(m_{\pi}^2)+\frac{m_K^2}{2}\nonumber \\
& &(2m_{\pi}^2-m_K^2)B(m_{\pi}^2,m_K^2,m_{\pi}^2)-(2m_{\pi}^4+m_K^4-
2m_{\pi}^2m_K^2-m_K^2p_1p_2)\nonumber\\
& &B_1(m_{\pi}^2,m_K^2,m_{\pi}^2)-2p_1p_2\Big(B_{00}(m_{\pi}^2,m_K^2,
m_{\pi}^2)+m_{\pi}^2B_{11}(m_{\pi}^2,m_K^2,m_{\pi}^2)\Big )\Big ]\nonumber\\
& &+\frac{1}{6(q^2+2qp_1+2qp_2)}\Big [-4m_{\pi}^2-3m_K^2+q^2+2qp_1+
2qp_2-4p_1p_2 \Big ] \nonumber \\
&&A(m_K^2)-\frac{1}{2(d-1)}A(m_K^2)+\frac{1}{4(d-1)}
(q^2-4m_K^2)B(q^2,m_K^2,m_K^2)\nonumber\\
& &+\frac{1}{q^2+2qp_1+2qp_2}\Big [(m_{\pi}^2+m_K^2+2p_1p_2)A(m_{\pi}^2)
+m_K^2(m_K^2-m_{\pi}^2)\nonumber \\
& &B(m_{\pi}^2,m_K^2,m_{\pi}^2)+2p_1p_2(m_K^2-m_{\pi}^2)
B_1(m_{\pi}^2,m_K^2,m_{\pi}^2)\Big]\nonumber \\
&&+2B_{00}(q^2,m_{\pi}^2,m_{\pi}^2)+2(m_K^2-m_{\pi}^2)C_{00}(m_{\pi}^2,
(p_1+q)^2,m_K^2,m_{\pi}^2)\nonumber\\
& &+\frac{1}{3(q^2+2qp_1+2qp_2)}\Big [4(2m_{\pi}^2-m_K^2+p_1p_2)A(m_{\pi}^2)
+2m_K^2(m_{\pi}^2\nonumber \\
&&+p_1p_2)B(m_{\pi}^2,m_K^2,m_{\pi}^2)-2[m_{\pi}^2(2m_{\pi}^2
-3m_K^2+2p_1p_2)+m_K^2(2m_K^2\nonumber \\
&&-p_1p_2)]B_1(m_{\pi}^2,m_K^2,m_{\pi}^2)\Big ]+\frac{2}{3}
\Big [(m_{\pi}^2-m_K^2+qp_1+p_1p_2)\nonumber \\
&&B((p_2+q)^2,m_K^2,m_{\pi}^2)+(m_K^2+q^2+2qp_1+
2qp_2+2p_1p_2)\nonumber \\
&&B_1((p_2+q)^2,m_K^2,m_{\pi}^2)\Big ]-\frac{4(3m_{\pi}^2-2m_K^2+p_1p_2)}
{3(q^2+2qp_1+2qp_2)}\Big [B_{00}(m_{\pi}^2,m_K^2,m_{\pi}^2)\nonumber \\
&&+m_{\pi}^2B_{11}(m_{\pi}^2,
m_K^2,m_{\pi}^2)\Big ]-\frac{1}{3}\Big
[(2m_{\pi}^2-3m_K^2+2qp_1+2p_1p_2)\nonumber \\
&&B(q^2,m_{\pi}^2,
m_{\pi}^2)+(3m_K^2+2q^2+4qp_1+4qp_2+4p_1p_2)\nonumber \\
&&B_1((p_2+q)^2,m_K^2,m_{\pi}^2)\Big ]-\frac{2}{3}
m_K^2(m_{\pi}^2-m_K^2+qp_1+p_1p_2)\nonumber \\
& &C(m_{\pi}^2,(p_2+q)^2,m_K^2,m_{\pi}^2)+\frac{1}{3}
\Big [4m_{\pi}^2(m_{\pi}^2-m_K^2+qp_1+p_1p_2)+m_K^4\Big ]\nonumber \\
& &C_1(m_{\pi}^2,(p_2+q)^2,m_K^2,
m_{\pi}^2)+\frac{1}{3}\Big
[4m_{\pi}^2(m_{\pi}^2-m_K^2+qp_1+qp_2+p_1p_2)\nonumber \\
& &+4qp_2(-m_K^2+
qp_1+p_1p_2)+m_K^4\Big
]C_2(m_{\pi}^2,(p_2+q)^2,m_K^2,m_{\pi}^2)-\frac{2}{3}
m_K^2\nonumber \\
&&\Big [C_{00}(m_{\pi}^2,(p_2+q)^2,m_K^2,m_{\pi}^2)+m_{\pi}^2 C_{11}(m_{\pi}^2,(p_2+q)^2,
m_K^2,m_{\pi}^2)\nonumber \\
& &+(m_{\pi}^2+qp_2)C_{22}(m_{\pi}^2,(p_2+q)^2,m_K^2,m_{\pi}^2)+
(2m_{\pi}^2+qp_2)\nonumber \\
&&C_{12}(m_{\pi}^2,(p_2+q)^2,m_K^2,m_{\pi}^2)\Big ]+\frac{2}{3}
B_{00}(q^2,m_K^2,m_K^2)\bigg \}.
\end{eqnarray}
\end{appendix}

\end{document}